# Reflectionless Programmable Signal Routers

Jérôme Sol[1], Ali Alhulaymi[2], A. Douglas Stone[2], Philipp del Hougne[3*]

[1] INSA Rennes, CNRS, IETR - UMR 6164, F-35000 Rennes, France

[2] Department of Applied Physics, Yale University, New Haven, Connecticut 06520, USA

[3] Univ Rennes, CNRS, IETR - UMR 6164, F-35000 Rennes, France

* Correspondence to philipp.del-hougne@univ-rennes1.fr.

## Abstract

We demonstrate experimentally that reflectionless scattering modes (RSMs), a generalized version of coherent perfect absorption, can be functionalized to perform reflectionless programmable signal routing. We achieve versatile programmability both in terms of operating frequencies and routing functionality with negligible reflection upon in-coupling, which avoids unwanted signal-power echoes in radio-frequency or photonic networks. We report *in-situ* observations of routing functionalities like wavelength demultiplexing, including cases where multi-channel excitation requires adapted coherent input wavefronts. All experiments are performed in the microwave domain based on the same irregularly shaped cavity with strong modal overlap that is massively parametrized by a 304-element programmable metasurface. RSMs in our highly overdamped multi-resonance transport problem are fundamentally intriguing because the simple critical-coupling picture for reflectionless excitation of isolated resonances fails spectacularly. We show in simulation that the distribution of damping rates of scattering singularities broadens under strong absorption so that weakly damped zeros can be tuned toward functionalized RSMs.

# Introduction

Signal routers, a pivotal ingredient of modern nanophotonic and radiofrequency (RF) networks aimed at distributing signals to transfer information or deliver energy, struggle to date with significant reflections upon signal injection. Besides the obvious loss in signal power, such reflections can give rise to devastating unwanted reflected-signal-power echoes in the network. In this article, we experimentally demonstrate that simultaneous perfectly reflectionless excitation of programmable signal routers is possible at multiple frequencies that can be injected through a single or multiple channels. Our concept leverages a complex scattering system with strong modal overlap that is massively parametrized by the hundreds of *in situ* tunable degrees of freedom of a programmable metasurface, to generate a scattering state that combines reflectionless excitation with the desired signal routing functionality. Based on the simple critical-coupling picture of balancing excitation and decay rates of an isolated resonance (*1–3*), our results are surprising because our system is strongly overdamped. However, because resonances are strongly overlapping in our system, as we explain below, differential absorption and eigenvalue repulsion can create some scattering singularities with much lower damping rates than the mean; these can be tailored to allow reflectionless excitation in combination with a desired routing functionality.

The reflectionless excitation of scattering systems is a fundamental wave-scattering problem across all areas of wave engineering that has received attention from diverse perspectives. *Perfect* reflection suppression has been known for decades to be possible through critical coupling. The latter, although often not clearly defined, typically refers to scenarios in which one seeks to excite a scattering system with a spectrally isolated resonance via a single channel (*1–3*). If the excitation and decay rate of this resonance can be exactly balanced at some frequency near the resonance frequency (assuming continuous wave (CW) excitation), there will be zero reflection. The critical-coupling concept typically does not distinguish between reversible and irreversible decay mechanisms, such as radiative "loss" vs. absorptive loss, respectively. If all decay mechanisms rest upon irreversible absorption, then incident waves are perfectly absorbed (*4, 5*). Critical coupling can also be understood in terms of the even older concept in electronics of perfectly matching an input impedance to a load (or output impedance) (*6*). However, the more general problem of reflectionless excitation of complex multi-resonance and/or multi-channel scattering systems received little attention until fairly recently because there was no general formalism which could precisely define the conditions for reflectionless excitation.

**Coherent Perfect Absorption.** Just over a decade ago, the concept of "Coherent Perfect Absorption" (CPA) defined the conditions for reflectionless excitations of arbitrarily complex linear scattering systems with a finite amount of loss (*7*), for the specific case of fully absorbed signals. CPA is defined based on the scattering matrix *S* which is a linear operator, depending on frequency, which maps input wave amplitudes in all scattering channels to output amplitudes in all scattering channels, i.e., it encodes the results of all possible scattering processes. For a finite scattering system, *S* can always be truncated to a finite $N \times N$ matrix. In our experiments, the asymptotic scattering channels are defined in an obvious manner through the *N* antennas coupled to the scattering structure, henceforth referred to as the cavity. The CPA theory describes generalized conditions under which an adapted input wavefront (the complex-valued amplitudes radiated by each of the *N* antennas, assumed to be mutually coherent) could be trapped and perfectly absorbed within the cavity. The requirement to observe CPA at some frequency is that *S* has a zero eigenvalue at that frequency and that the corresponding eigenvector is used as input wavefront – irrespective of the complexity of the cavity (e.g., isolated vs. overlapping resonances). Discrete, countably infinite solutions of this type ("zeros" of the *S*-matrix (*8, 9*)) exist if the frequency, $\omega$, is defined in the complex plane, but only for a real $\omega$ solution is there a reflectionless CW input wavefront at $\omega$. Clearly the CPA condition can only be achieved if the cavity has internal loss, describing absorption within the cavity and transduction of the energy to other degrees of freedom, e.g., heat or electricity.

To achieve CPA in a cavity with loss, one must in general tune a parameter so that one of the complex $\omega$ solutions ends up lying exactly on the real axis at some $\omega_0$; then, if we input the correct wavefront at $\omega_0$ it will be perfectly absorbed. Initially, CPA was mainly studied in relatively simple cavity geometries with two channels (*10*), and the ability to interferometrically control light with light (rather than through nonlinear effects) led to proposals of CPA-based signal modulators (*11–15*). In the past few years, there have been a number of theoretical (*16–19*) and experimental (*20, 21*) studies of the CPA effect in complex scattering geometries, often with many scattering channels. Importantly for the current work, some studies in similar systems to the current one have gone well beyond just studying such unconstrained CPA and have optimized perfectly absorbing cavities under various constraints (*22–26*). To achieve certain CPA-based functionalities like secure information transmission (*22*) and analog computation (*23*), it is important to achieve CPA at a specific rather than arbitrary real frequency. However, there is no guarantee that such constraints can be successfully imposed simply by tuning a single or few parameters of a (complex) scattering system. Heuristic insights from recent experiments show that the combination of a cavity with strong modal overlap (i.e., a high density of zeros in the complex plane) with a massive parametrization (e.g., hundreds of degrees of freedom from a programmable metasurface) makes it possible to precisely impose such constraints in an *in situ* reprogrammable manner (*22*, *23*). A programmable metasurface is an ultrathin array of elements termed "meta-atoms" whose scattering properties can be individually reconfigured (*27*); inside a wave-chaotic cavity, a programmable metasurface enables the tailoring of the boundary conditions (*28*). Nonetheless, while constrained CPA can be applied to achieve some functions, it is not suitable for the signal-routing functions studied here, precisely because CPA is defined by perfect absorption of the signal. However, a more recent generalization of the CPA concept, termed "reflectionless scattering modes" (RSMs) (*29*), provides exactly the necessary framework for understanding reflectionless signal routing in complex, tunable cavities.

**Reflectionless Scattering Modes (RSMs).** The RSM theory defines the conditions for reflectionless excitation of an arbitrarily complex linear scattering systems (with or without internal loss) through $N_{\text{in}} \leq N$ of the $N$ scattering channels (the case $N_{\text{in}} = N$ corresponds to CPA and will not be our focus below). Since our aim is to route signals "forward" into particular output ports, each functionality will define a set of $N_{\text{in}}$ input ports, where the signals enter, and a set of $N - N_{\text{in}}$ output ports, where the signals exit, with zero reflection back into the input ports. RSM theory constructs an $N_{\text{in}} \times N_{\text{in}}$ sub-matrix of *S*, denoted as $R_{\text{in}}$, and searches for its eigenvectors with eigenvalue zero. Like in CPA, typically a countably infinite set of solutions exist in the complex $\omega$ plane, constituting a different complex spectrum than the *S*-matrix zeros (*30*); we refer to these new solutions as "*R*-zeros" (*29*, *31*). As noted, only if an *R*-zero lies on the real axis, does there exist a CW adapted wavefront for reflectionless excitation through the chosen $N_{\text{in}}$ channels. We refer specifically to such real $\omega$ solutions as RSMs. The incident signal energy of an RSM is partially or fully scattered into the output ports, depending on whether the cavity has internal loss or not.

Precursors of the RSM concept already highlighted that special symmetries, such as $\mathcal{PT}$-symmetry, can drastically improve the accessibility of RSMs (*32*, *30*). For general systems without such symmetries, tuning is necessary to ensure that one of the *R*-zeros lies somewhere on the real axis and becomes an RSM. However, zero reflection at a single arbitrary frequency alone is rarely a desired functionality, if there are multiple output ports; instead, one would like to steer signals at specific frequencies in specific ways, typically simultaneously at multiple frequencies, e.g., for demultiplexing. Because this is a linear theory, one can aim to optimize the system simultaneously for different target behaviors at distinct frequencies. We show in the current work that a massively parametrized metasurface-programmable cavity with strong modal overlap allows us to build desirable routing functionalities upon RSMs, something not shown previously, either theoretically or experimentally. As noted, even finding RSMs in our highly overdamped cavity is surprising and cannot be explained by the critical-coupling concept. Hence, we have also performed substantial numerical simulations in this regime of overlapping resonances to confirm that RSMs exist in such overdamped systems.

**Signal Routing in Nanophotonics and Acoustics.** Independent of this strand of research on perfectly reflectionless excitations under complex scattering, nanophotonics and acoustics research groups have set out to design signal-routing devices. A prototypical example are wavelength demultiplexers that split an input signal composed of two superposed frequencies into two single-frequency outputs. For a fixed pair of operating frequencies, topology-optimized designs of refractive-index patterns in integrated photonic devices have been proposed to achieve this signal-routing functionality (*33, 34*). However, once fabricated, the operation of these devices is fixed to a specific frequency pair and signal-routing functionality, and they have the drawback of not being based on any general or intuitive understanding. A significant advance for programmable signal routing was made in Ref. (*35*) which projected a light pattern onto a multi-mode waveguide to locally alter the refractive index and thereby imprint a desired scattering topology. The projected patterns were massively parametrized by a digital micromirror device and could be reprogrammed to route a single-frequency input to a desired output channel. This work emphasized the advantage of the massive parametrization to provide enough tunable degrees of freedom, in contrast to tuning with a few thermal or electro-optic elements (*36–39*), and is similar in spirit to the massive programmable-metasurface parametrization used in the current work. Encouragingly, the propagation loss in the massively parametrized multi-mode waveguide was very low. Unlike our metasurface-parametrized cavity, the parametrization of the waveguide was seemingly linear without significant multiple scattering or back-reflection. However, the reported routing functionalities were limited to a single frequency. Moreover, for all of the here-discussed static and programmable nanophotonic signal routers, low reflection upon injection (low insertion loss) was, of course, a design goal, but the proposed designs did not achieve very high reflection suppression, nor did they establish the conceptual link to the theory of reflectionless scattering.

Meanwhile, sub-wavelength resonant acoustic structures were used to perform wave-routing functionalities that were controlled by an adapted input wavefront (*40–42*), following the above-mentioned idea of "controlling light with light" (*11–15*). These works differ from ours chiefly in that the input wavefront rather than the scattering system determines the wave-routing functionality, relegating the functionality control to wave sources that must offer multi-channel coherent control. Moreover, these systems can only implement specific rather than arbitrary routing functionalities, at a single frequency. While their operation is reflectionless in the ideal structures treated in analytical calculations, corresponding experiments struggle to achieve *very high* reflection suppression because these systems have only limited or zero capability of *in-situ* corrections for fabrication inaccuracies.

**RSM-based Signal Routing.** In this paper, we experimentally demonstrate reflectionless programmable signal routers based on a wave-chaotic scattering system that is massively parametrized by a programmable metasurface. Our system's high modal density combined with the hundreds of tunable degrees of freedom allows us with unprecedented precision to impose simultaneously deep suppression of all reflections and of all undesired transmissions at multiple arbitrary desired frequencies, while also generating maxima in the transmission into the desired ports. Moreover, our system's inherent programmability allows us to toggle *in situ* between configurations optimized for different pairs of operating frequencies and signal-routing functionalities. We experimentally confirm the perfectly reflectionless routing with *in-situ* observations of the power exiting our system through all connected channels upon direct simultaneous excitation at two frequencies through one or multiple channels.

The fundamental mechanism of the reflection suppression is a complicated, tunable destructive interference at the input ports arising from the multiple scattering of the input wavefront within the cavity. Prior to this work, there was no significant study in either theory or experiment of whether such reflectionless programmable routing was possible. In fact, there was not even a clear theoretical understanding of whether merely observing RSMs (let alone the signal routing functionality) is possible for a highly overdamped cavity, where the internal cavity damping far exceeds the total output coupling. We show below that the simple critical-coupling picture fails under these conditions and our simulations of the $R$-zero spectrum of a model for a cavity similar to ours give insight into why RSMs exist. While the mean imaginary part of the spectrum of $R$-zeros is very large (and negative) for a highly overdamped cavity, the distribution of $R$-zeros broadens in the overdamped regime, making it relatively easy to tune zeros very close to the real axis. The universality of the analytic properties of scattering operators implies that our approach to reflectionless programmable signal routing can be applied to other complex scattering systems (multi-mode waveguides, quantum graphs, etc.), to other techniques for massive parametrization, and other wave phenomena (acoustics, elastics, optics, quantum mechanics).

## Results

**Unconstrained RSMs in Strongly Overdamped Systems.** To start, we examine the possibility of observing unconstrained RSMs in our experimental setup depicted in Fig. 1B. This is a precondition for being able to program additional output functionality into the system, and is not trivial from the theoretical perspective, as discussed above. We also want to confirm that the qualitative behavior is consistent with the general RSM theory. Our scattering system consists of an irregularly shaped three-dimensional metallic enclosure that is parametrized by 304 1-bit programmable meta-atoms (see Methods and Supplementary Note 1 for details). Four single-mode guided channels are coupled to the scattering system via coax-to-waveguide adapters. The scattering-matrix formalism is illustrated for the example of an RSM involving $N_{\text{in}} = 2$ out of $N = 4$ channels (the ones indexed 1 and 2) in Fig. 1A. As illustrated there, a wavefront $[a_1\ a_2]$ is injected through the channels indexed 1 and 2 and ideally we impose that zero energy exits the system through these two channels, implying that the submatrix $R_{\text{in}}$ of $S$ involving the two injection channels, highlighted in green, has a zero eigenvalue with corresponding eigenvector $[a_1\ a_2]$. In general, given the absence of any special symmetries in our system (e.g., $\mathcal{PT}$-symmetry), it is unlikely that there would be an $R$-zero sufficiently close to the real-frequency axis to generate a reflection dip that is many orders of magnitude below the background for a given random configuration of the metasurface (*22*). In non-symmetric systems, tuning is generically necessary to move a reflection zero to the real frequency axis (*29, 31*). (Note that the condition of zero imaginary part will never be exactly satisfied due to experimental limitations, so in the experimental context we use the term "reflectionless" to refer to a deep dip below the background reflection, as defined below). Moreover, there is strong non-localized surface absorption of microwaves in our metallic system, implying that $S$ is highly sub-unitary: the enclosure's composite quality factor is 369 and the average transmission magnitude between two ports is $-28.4$ dB, implying that the internal loss is orders of magnitude larger than the total radiative coupling in or out of the cavity. In the simplest model of the effect of absorption on the $R$-zeros, the distribution of zeros is simply translated downward in the complex frequency plane by the internal loss rate. If this were the case, as we discuss further below, all $R$-zeros would have a large imaginary component with negative sign, leading to high reflection of CW input waves, due to the overdamping impedance mismatch, and tuning would be unable to bring a zero to the real axis.

However, upon scanning a large parameter space of $10^4$ random metasurface configurations and 1601 frequency points between 4.9 GHz and 5.6 GHz with the stringent criterion for an RSM of below $-65$ dB of reflected power, we do find a considerable number of RSMs. Recall that, since we have four ports, $2^4 - 1 = 15$ different reflectionless boundary conditions are possible, which can be grouped according to the number of input ports, $N_{in}$, with $N_{in} = 4$ corresponding to CPA. Varying $N_{in}$ allows us to check a qualitative prediction of the RSM theory. On average, the larger is $N_{in}$, the higher in the complex plane the R-zeros lie; for instance, the zeros of the full S-matrix should on average have a larger imaginary part than other R-zeros. Hence, once we include the strong absorption damping, which moves almost all zeros below the real axis, the distribution of zeros of the full S-matrix should be closer to the real axis, making it easier to tune them to the real axis (CPA), than the R-zeros with $N_{in} = 3,2,1$. Similarly, we would expect to find more RSMs for $N_{in} = 3$, than for $N_{in} = 2$, and so on. This is precisely what we observe in Fig. 1C: the unconstrained RSMs are indeed more likely to arise (with tuning) for larger $N_{in}$. This gives us confidence in interpreting our results with the RSM theory.

Our explanation for the existence of RSMs in a highly overdamped cavity and for the trend in Fig. 1C involves the distribution of R-zeros for different values of $N_{in}$ and damping, which is not a quantity accessible in experiments. Hence, we study this distribution through numerical simulations of a two-dimensional chaotic cavity with three ports, as shown in the inset of Fig. 1D. Finding the complex R-zero spectrum involves essentially the same techniques as finding the resonance spectrum of a complex cavity, and is efficiently done in this case by a variant of the PML method – see Methods and Supplementary Note 5 for details. Using this approach, we identify all R-zeros for the different possible choices of injection through $N_{in} = 3, 2,$ or 1 channel(s) within a large frequency interval, as well as the poles (resonances) for which $N_{in} = 0$ and $N_{out} = 3$. We then generate statistical distributions of their imaginary parts and calculate their mean and variance. In our simulations, we can conveniently perform this analysis for different ratios of absorption rate to radiation coupling, $|\Gamma_{abs}/\Gamma_{rad}|$, by adding a finite conductivity to the walls (28, 43) and varying it appropriately (see Supplementary Note 5).

We plot in Fig. 1D the mean and standard deviation of each of these distributions of imaginary parts for different values of $N_{in}$ as a function of $|\Gamma_{abs}/\Gamma_{rad}|$, and we also plot the largest imaginary part observed for R-zeros with $N_{in} = 3$ (zeros of the full S-matrix). Our results yield three important conclusions: first, as anticipated, the mean of the distributions of imaginary components decreases monotonically as $|\Gamma_{abs}/\Gamma_{rad}|$ is increased for all considered quantities. Second, for all values of $|\Gamma_{abs}/\Gamma_{rad}|$, the means are strictly ordered and increasing for larger $N_{in}$, as expected from the RSM theory. Third, the standard deviations of the distributions of imaginary components increase monotonically as $|\Gamma_{abs}/\Gamma_{rad}|$ is increased. If this had not been the case, one can infer from the plot that for the largest values of $|\Gamma_{abs}/\Gamma_{rad}|$ (similar to our experimental system) even the least damped CPA state would be several standard deviations away from the real axis, making it extremely difficult to achieve CPA (or RSM) with tuning. However, the broadening of the distributions with increasing $|\Gamma_{abs}/\Gamma_{rad}|$ implies that despite strong overdamping, some zeros can be found in the vicinity of the real frequency axis, making tuning to find RSMs feasible. For the largest damping shown in Fig. 1D, the extremal zero for CPA is only 0.66 standard deviations away from the real axis.

One source of this broadening is due to differential absorption. Simulations detailed in Supplementary Note 5 show that some $R$-zeros are more sensitive to absorption than others, leading to differential spreading of the $R$-zero distributions along the imaginary frequency axis. The most sensitive $R$-zeros correspond to the more localized eigenmodes, which also have higher overlap with the walls. These partially localized eigenmodes may have a semiclassical origin in terms of scars or stable periodic orbits, as discussed in Supplementary Note 5. Another potential source of broadening is enhanced eigenvalue repulsion as the resonances overlap more. In systems with only scattering loss, a spreading of the resonance distribution along the imaginary frequency axis has been found as the outcoupling is increased and is associated with an effect referred to as "resonance trapping" (*44*, *45*). Similar studies of the effect of absorption loss on complex eigenvalue distributions have not been done, to our knowledge. The simulations of zero motion in the Supplementary Note 5 do show some apparent interactions of eigenvalues, but not the dramatic effects found in resonance trapping models. Hence, more work is required to confirm and quantify the increase of eigenvalue repulsion for a strongly absorption-damped wave system such as ours.

Note that in our simulations we have not performed an optimization of the cavity geometry to achieve RSMs, and so the extremal values shown do not present an upper bound on the achievable imaginary component in an optimized system. Our initial studies with geometric tuning indicate that it is straightforward to generate RSMs with $N_{in}$ = 3, 2, or 1 with optimization in the overdamped regime; an example with $N_{in}$ = 1 is shown as inset in Fig. 1D.

So far, we have reported the first rigorous experimental observation of unconstrained RSMs in overmoded scattering systems without symmetry and provided insight into the underlying distributions that enable these observations which cannot be explained in terms of the simple critical-coupling picture. However, as noted above, such unconstrained RSMs are not yet well aligned with practical technological needs. Thus, before closing this section, as a first step toward constrained RSMs of practical value, we consider the constraint of generating RSMs at a specific frequency and choice of input channels; we choose 5.2 GHz as the frequency, and injection through the channels indexed 1 and 2 in our experimental setup. The closest RSM from our unconstrained search underlying Fig. 1C is plotted in purple in Fig. 1E and does not yield any reflection suppression at the desired frequency. However, following the approach to on-demand CPA introduced in Ref. (*22*), we can optimize the metasurface configuration to impose the desired RSM on our scattering system. To that end, we perform an iterative optimization of the metasurface configuration, as seen in Fig. 1F and detailed in the Methods and Supplementary Note 3. Thereby, we eventually identify a system configuration for which we have the desired {1,2}-RSM at exactly 5.2 GHz. Because we consider multi-channel excitation, the reflection is only suppressed if the correct adapted wavefront (found as part of the optimization and shown in Fig. 1G) is injected. In addition to the reflected power, we also plot in Fig. 1E the power transmitted into the two remaining ports indexed 3 and 4. It is apparent that the RSM does not coincide with a zero or maximum in any of these transmissions. As stated above, frequency-constrained RSMs are hence a necessary but not sufficient condition for reflectionless signal routing.

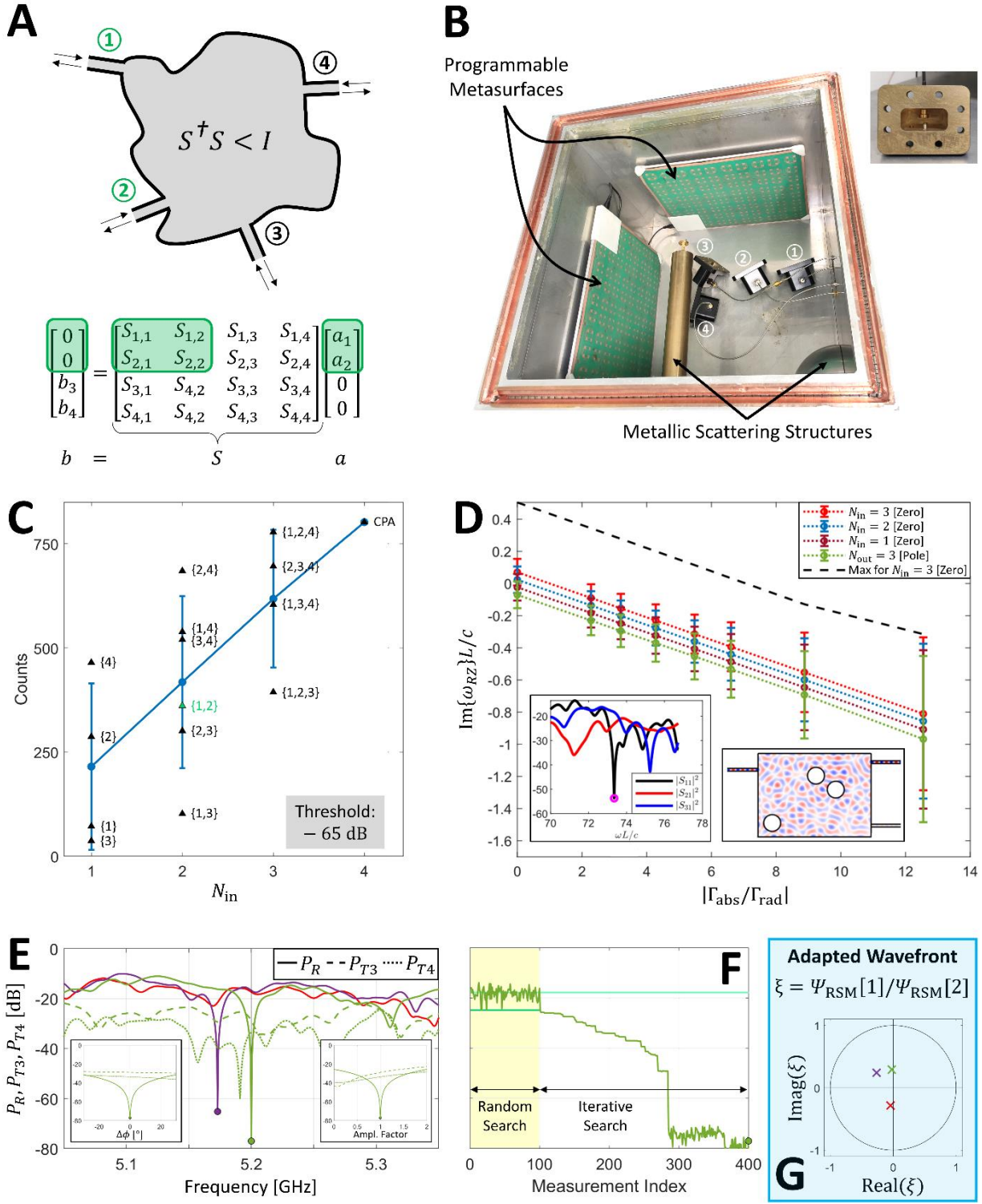

**Fig. 1. Concept and experimental observation of RSMs without and with frequency constraint in a metasurface-programmable overmoded lossy scattering system.** (**A**) Generic schematic of a non-unitary scattering system with four attached channels, and matrix formalism corresponding to an RSM involving channels 1 and 2 (highlighted in green). (**B**) Photographic image of the corresponding experimental setup comprising a metallic electrically large scattering enclosure with irregularly shaped metallic scattering structures (top cover removed to show interior), two programmable metasurfaces composed of 152 meta-atoms each, and four waveguide-to-coax adapters to couple four monomodal channels to the system. The inset shows the front view of a waveguide-to-coax adapter. (**C**) Number of RSMs found

across $10^4$ random metasurface configurations and 1601 frequency points between 4.9 GHz and 5.6 GHz. Black triangles indicate counts for specific indicated choices of $N_{in}$ input ports, blue dots and errorbars represent the average and standard deviation across all possible choices. The labels indicate the chosen $N_{in}$ injection channels. (**D**) Numerically determined mean and standard deviation of the distributions of the imaginary components of zeros and poles (color-coded for various choices of $N_{in}$ and $N_{out}$) as a function of the ratio of absorption coupling to radiative coupling. The distributions are based on the PML method applied within a certain frequency interval to the cavity depicted in the inset (see Supplementary Note 5). In addition, the black-dashed line indicates the highest imaginary component for $N_{in} = 3$ found in the interval. (**E**) Reflected signal power $P_R$ upon excitation through channels 1 and 2 for a random metasurface configuration (red), for the RSM closest to 5.2 GHz out of the data from (C) (purple), and for a metasurface configuration optimized for 5.2 GHz. The dashed and dotted lines present the powers $P_{T3}$ and $P_{T4}$ transmitted into channels 3 and 4, respectively. The insets show the sensitivity of the RSM optimized with frequency constraint to detuning of the relative phase (left, detuned by $\Delta\phi$) or relative amplitude (right, detuned by a multiplicative scaling factor) of the two RSM wavefront $\Psi_{RSM}$ entries. (**F**) Optimization dynamics of the frequency-constrained RSM from (E): first, 100 random metasurface configurations are tested; second, the best configuration is iteratively optimized further. (**G**) The two-channel input wavefronts corresponding to the curves in (E) are visualized in terms of the relative phase and amplitude difference using the same color codes as in (E).

**Reflectionless Wavelength Demultiplexer.** Our first goal is the implementation of a 3-port *reflectionless* wavelength demultiplexer: two frequencies $f_1$ and $f_2$ are injected without any reflection through Port 1, with $f_1$ transmitted to Port 3 but not Port 2, and $f_2$ transmitted to Port 2 but not Port 3 (see Fig. 2A). This device thus involves six constraints that must be satisfied simultaneously, a major step beyond the frequency-constrained RSM shown in Fig. 1E, with only a single constraint. First, two simultaneous frequency-constrained RSMs with $N_{in} = 1$ must be imposed. Here, we deliberately focus on single-channel input excitation because of its higher technological relevance (no need for very costly coherently controlled multi-channel sources); however, a case of signal routing with multi-channel excitation is presented in the last section of this paper. Note again that due to the strong modal overlap these single-channel RSMs cannot be understood in terms of the simple critical-coupling picture. Second, two simultaneous frequency-constrained zeros in undesired transmission must be imposed. Third, two simultaneous frequency-constrained maxima in desired transmission must be imposed. Note that constraints on transmission properties are beyond the RSM framework, which only demands zero reflection, and the general theory implying the robust existence of *R*-zeros, and RSMs with tuning, does not guarantee that such solutions exist.

For this three-port experiment, we remove one waveguide-to-coax adapter from the setup seen in Fig. 1B. Because the cost function to be minimized must trade off all six constraints, we heuristically identify a suitable weighting of the six constraints as detailed in the Methods and Supplementary Note 3. For the experimental studies above, the functionality of interest could be extracted from measuring the *S*-matrix on a dense grid of frequencies. Here, while we will use this approach to *find* optimal metasurface settings, we will demonstrate reflectionless wavelength demultiplexing *in situ* for our system, by actually inputting the two chosen frequencies simultaneously, and measuring the outputs on all channels. For this purpose, we implement a versatile measurement apparatus that can be switched between two measurement modalities (see Fig. 2B). In the first modality, we can measure our system's full scattering matrix with a vector network analyzer; in the second modality, we can inject *in situ* the sum of two continuous-wave (CW) signals at $f_1$ and $f_2$ and measure the spectrum of the signals exiting the scattering system through each connected channel (including the injection channel) on a spectrum analyzer. Further details about the measurement apparatus are provided in the Methods and Supplementary Note 1.

The frequency-dependent scattering matrix of our system, optimized to demultiplex the two frequencies 5.1 GHz and 5.2 GHz, is shown in Fig. 2C, Fig. 2E, and Fig. 2G. In Fig. 2C, the sharp reflection dips characteristic of RSMs are seen exactly at the two desired frequencies. Similarly, the dips in undesired transmissions are found to be quite narrow whereas the maxima in desired transmissions are rather broad. We conjecture that this difference in broadness originates from the fact that the transmission zeros can also be defined as spectral singularities (zeros of a scattering coefficient), although no details have been worked out for this, whereas the transmission maxima are not. The zero-scattering regime, be it for perfect suppression of reflection or undesired transmission, has intriguing effects on the delay of wave propagation through the system. Indeed, zeros of scattering coefficients that lie on the real frequency axis lead to phase singularities and anomalously-long diverging delay times (*25*, *46–50*). The interpretation and statistical properties of complex Wigner time delays in sub-unitary (and possibly overmoded) scattering systems, and their relation to the singularities (poles and zeros) of the associated wave-transport matrix, is currently an active area of research (*26*, *49–52*). In general, away from such singularities, it is well established that longer dwell times increase the sensitivity to minute perturbations, with important implications for precision sensing (*53*, *54*). Hence, zero-scattering with diverging dwell times may imply an extreme sensitivity (*22*, *25*, *48*) – provided that the waves are indeed infinitely trapped and bouncing around without being absorbed or radiated away. We certainly observe an extreme sensitivity to frequency detuning both for zeros in reflection (guaranteed to exist by analytic properties of the *S*-matrix) and undesired transmission.

Based on the corresponding *in-situ* observations, we determine the performance metrics of our reflectionless wavelength demultiplexer. To eliminate signal loss in the measurement apparatus that is not related to the scattering system, we perform a calibration measurement without the scattering system (see Methods and Supplementary Note 2). The difference between *in-situ* measurements with the optimized scattering system and the calibration measurements directly yields the performance metrics. As seen in Fig. 2D, we achieve a reflection suppression of at least −59 dB for both frequencies, justifying the terminology "reflectionless" for our signal router. At the same time, Fig. 2F and Fig. 2H reveal that undesired transmissions are suppressed by at least 60 dB whereas desired transmissions are attenuated by at most 20 dB, yielding a very strong discrimination between desired and undesired transmission of 40 dB in our demultiplexer.

The 20 dB attenuation of desired transmission is not an inherent limitation of our concept but specific to the structure of our proof-of-principle experiment in which we employ a strongly absorbing microwave cavity, which causes an average transmission attenuation of 28.4 dB. Given this baseline transmission attenuation, it is striking that our optimized system yields more than 8 dB of improvement on desired transmission. Ideally, of course, the attenuation of desired transmission would be negligible; simulations of lossless cavities show that optimization does still yield a good demultiplexing structure (see Supplementary Note 5). This goal seems certainly within reach in practice, if our concept is transposed to other scattering systems with much less overall attenuation, such as the nanophotonic silicon-on-insulator device from Ref. (*35*) with almost no propagation loss. It should also be noted that some desired-transmission attenuation may be tolerable in exchange for perfect reflection suppression in scenarios where the signal router is part of a network that is vulnerable to reflected-power echoes. For example, in modern high-frequency RF transceiver chains, especially active components like amplifiers are at risk of becoming unstable due to reflected-power echoes. Ultimately, these effects can result in a complete malfunction of the entire front-end chain, which is why currently reflected-power echoes are avoided by inserting additional isolators, circulators or in-line attenuators into the chain. Isolators or circulators are bulky, costly and may involve power-consuming nonreciprocal elements like transistors. Attenuators are reciprocal and hence also attenuate the desired signal. Similarly to these considerations for RF networks, reflected-power echoes negatively impact the stability of laser sources and other components in integrated photonic networks (*55*). Thus, reflectionless signal routers may be technologically attractive even if they involve attenuation of the desired transmissions because they remove the need for additional components that are currently used to suppress the consequences of reflected signals.

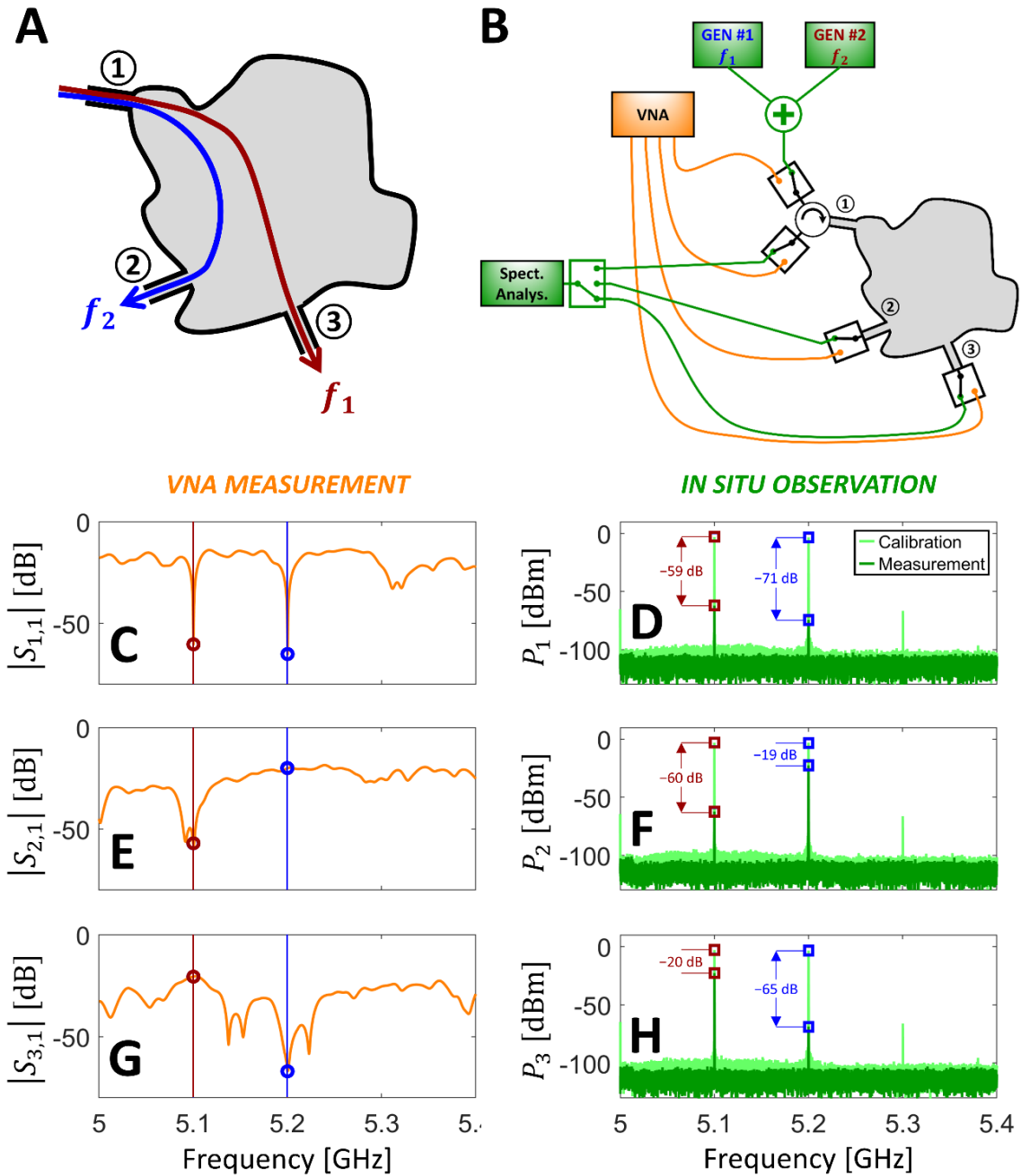

**Fig. 2. Reflectionless wavelength demultiplexer.** (**A**) Schematic of targeted wavelength demultiplexing functionality. (**B**) Schematic drawing of experimental setup for VNA measurements and *in situ* observation. (**C**,**E**,**G**) VNA measurements of the scattering parameters of the optimized system. (**D**,**F**,**H**) *In situ* measurements of the optimized system. Continuous-wave signals at $f_1$ and $f_2$ are summed and injected through channel 1 [see (B)]. The power exiting the system through all three channels is measured and plotted in dark green. Additionally, a calibration measurement to characterize losses in the cables of the measurement setup [see (B)] is shown in light green. The difference between the two curves corresponds to the *in situ* observed performance: reflection of at most −59 dB, undesired transmission of at most −60 dB, desired transmission of at least −20 dB, and hence transmission discrimination of 40 dB.

**Programmability of Reflectionless Signal Routers.** Having implemented a specific reflectionless signal router (specific choice of operating frequencies, specific choice of routing functionality) in the previous section, we now turn our attention to the *in-situ* reprogrammability of our system that is inherently offered by our approach. Especially for today's frequency-agile RF systems (e.g., cognitive radio), the ability to reprogram a signal router *in situ* so that it operates with perfect reflection suppression for a different pair of frequencies is important and valuable. We demonstrate this ability in Fig. 3A where we show *in-situ* observations of reflectionless wavelength demultiplexing for four different pairs of operating frequencies. There is, of course, no special relation between the chosen frequencies and the geometry of our scattering system, and, in principle, arbitrary frequencies could be chosen as long as they are not very close and fall into the roughly 400 MHz interval around 5.15 GHz in which our programmable metasurface can efficiently manipulate the field (see Supplementary Note 1). If frequency agility over an even wider range is desired in the future, alternative metasurface designs or metasurfaces composed of meta-atoms operating within multiple bands can be deployed.

The *in-situ* measured performance metrics of the reflectionless wavelength demultiplexing results from Fig. 3A are summarized in Table 1. For all considered choices of operating frequencies, we achieve at least −50 dB of reflection suppression and at least 34 dB in transmission discrimination. These metrics can be further improved in the future by improving the parametrization of the scattering system, e.g., through even more programmable meta-atoms, through multi-bit programmability of the programmable meta-atoms, etc. In addition, we also display the corresponding optimized metasurface configurations in Fig. 3A. The non-intuitive nature of these configurations due to the complex nature of our scattering system is apparent. There is also no obvious relation between the optimized patterns for different choices of operating frequencies. The different programmable meta-atom states are statistically roughly equally represented.

| $f_1$ [GHz] | $f_2$ [GHz] | $P_R$ [dB] | $P_{Tu}$ [dB] | $P_{Td}$ [dB] | D = $P_{Td}$ - $P_{Tu}$ |
|---|---|---|---|---|---|
| 5.1 | 5.2 | −59 | −60 | −20 | 40 |
| 5.1 | 5.3 | −51 | −60 | −21 | 39 |
| 5.05 | 5.35 | −50 | −55 | −21 | 34 |
| 5.2 | 5.3 | −52 | −54 | −20 | 34 |

**Table 1. Performance metrics of the *in-situ*-observed demultiplexer functionalities displayed in Fig. 3A.** *$P_R$, $P_{Tu}$ and $P_{Td}$ denote reflected, undesirably transmitted and desirably transmitted*

**power.**

Another important reconfigurability feature of our system concerns the signal-routing functionality. So far, we focused on wavelength demultiplexing. But by suitably redefining the constraints, our system can also be optimized for any other input-output functionality. We demonstrate this feature in Fig. 3B for an unconventional routing functionality for which $f_2$ is injected through the port through which $f_1$ exits, and $f_2$ exits through the remaining third port. For this routing functionality, simultaneous RSMs of $f_1$ and $f_2$ must be imposed on distinct filtered scattering matrices. The modified measurement apparatus is detailed in Supplementary Note 1. The *in-situ* observations displayed in Fig. 3B, and the corresponding performance metrics summarized in Table 2 (at least 52 dB of reflection suppression and at least 36 dB of transmission discrimination), confirm that both operating frequencies and routing functionality can be reprogrammed *in situ* while guaranteeing perfect reflection suppression.

| $f_1$ [GHz] | $f_2$ [GHz] | $P_R$ [dB] | $P_{Tu}$ [dB] | $P_{Td}$ [dB] | D = $P_{Td}$ - $P_{Tu}$ |
|---|---|---|---|---|---|
| 5.1 | 5.2 | −53 | −57 | −20 | 37 |
| 5.1 | 5.3 | −55 | −57 | −21 | 36 |
| 5.1 | 5.15 | −52 | −61 | −21 | 40 |
| 5.15 | 5.2 | −52 | −64 | −21 | 43 |

**Table 2. Performance metrics of the *in-situ*-observed routing functionalities displayed in Fig. 3B.**

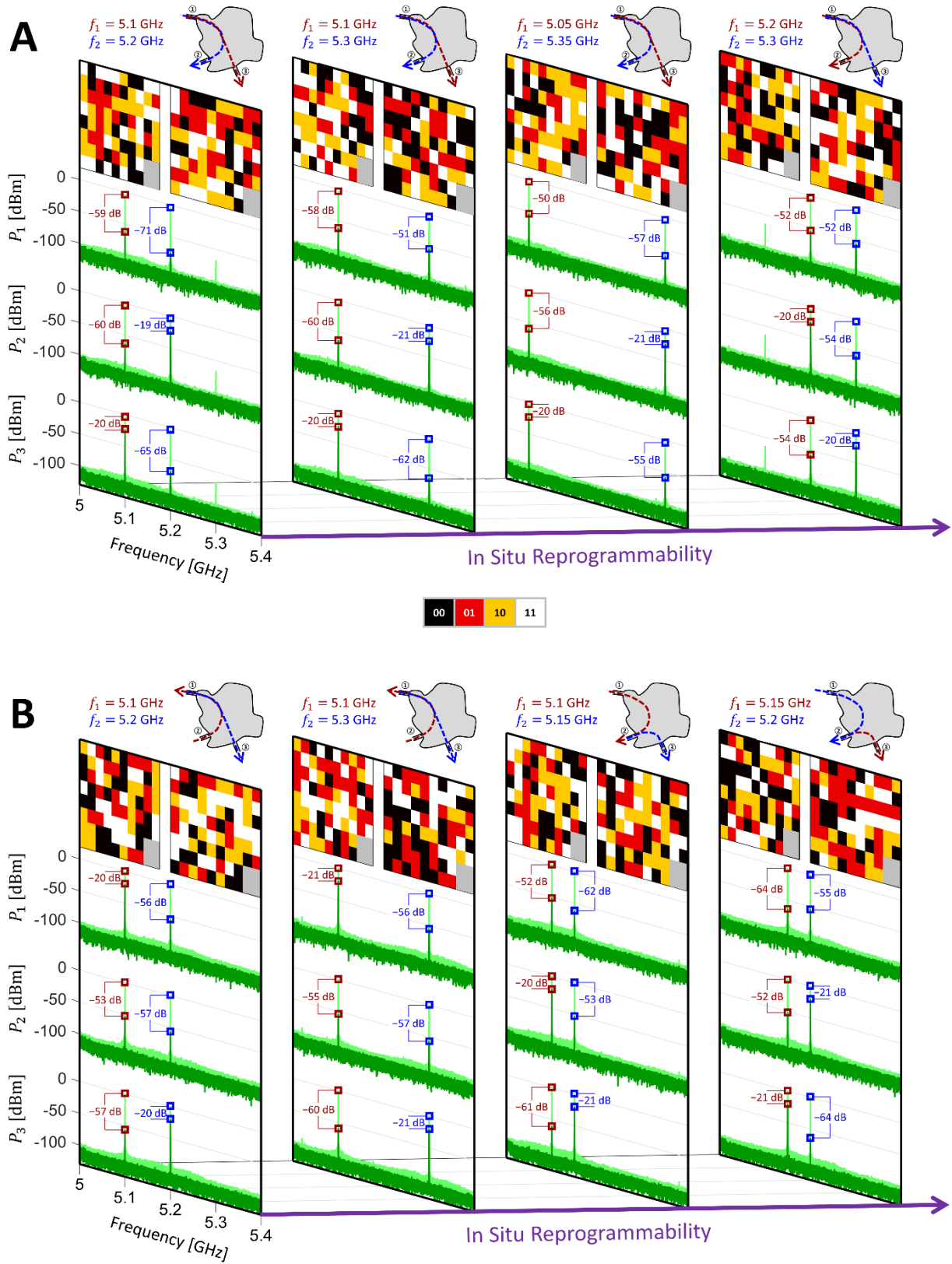

**Fig. 3. Reprogrammability of reflectionless signal routing frequencies and functionality.** (A) *In situ* observations of four instances of reprogramming *in situ* the wavelength demultiplexing frequencies by reconfiguring the programmable metasurface are shown. The first instance is that from Fig. 2. The corresponding metasurface configurations are indicated; the color code identified in the legend encodes the 1-bit programmable configuration of the meta-atom with respect to the two orthogonal field polarizations. (**B**) *In situ* observations of a different signal routing functionality by reprogramming the metasurface configuration. Here, $f_2$ is injected via the channel through which $f_1$ is supposed to exit, and $f_2$ is desired to exit through the remaining third channel. Again, four instances for different choices of $f_1$ and $f_2$ are shown.

**Multi-Channel Excitation with Adapted Wavefront.** To this point, we have considered routers in which each frequency is injected only through a single channel; in this section, we consider the case of multi-channel excitation of one of the injected frequencies. Because inputs at the same frequency from different channels will interfere, only a special adapted wavefront can be injected without reflection, as discussed previously. This multi-channel case further manifests the generality of our approach, especially since much of the recent academic interest in CPA and related phenomena focused on this need for a non-trivial adapted wavefront. From a technological perspective, however, the generation of an adapted wavefront is very costly and hence unattractive. Coherent multi-channel wave control requires individual phase and amplitude modulation of each source, as well as synchronization of the sources. It is worth noting that the previously discussed single-channel excitation cases are already highly non-trivial and intriguing from an academic perspective due to the strong modal overlap that leads to a failure of simple critical-coupling intuition.

To study a multi-channel-excitation scenario, we focus on yet a different routing functionality: wavelength multiplexing. Note that time-reversing a reflectionless wavelength demultiplexer does not yield a wavelength multiplexer due to the presence of absorption, which breaks the time-reversal symmetry of the *S*-matrix. Specifically, we now consider the 4-port setup from Fig. 1B. The targeted device is excited with $f_1$ through Ports 1 and 2 using wavefront $\psi_{\text{in}}$, and with $f_2$ through Port 3, and energy at both frequencies exits the system only through Port 4. The modified measurement apparatus is detailed in Supplementary Note 1. The reflected power at $f_1$ is now defined as the sum of the powers exiting the system through the two injection channels upon excitation with the wavefront $\psi_{\text{in}}$ that coherently minimizes the reflection. Analogous to coherent enhanced absorption (CEA) (*56*), coherently minimized reflection (CMR) is obtained by defining $\psi_{\text{in}}$ as the eigenvector corresponding to the smallest eigenvalue of $R^\dagger R$, where $R$ now is a filtered scattering matrix involving the channels indexed 1 and 2. CMR does not distinguish between absorption and radiation "loss" whereas CEA assumes that all loss is of absorptive nature. If the smallest eigenvalue of $R^\dagger R$ is zero, then the CMR wavefront coincides with the RSM wavefront (just as the CEA wavefront would coincide with the CPA wavefront in the case of a zero eigenvalue). Moreover, the undesirably transmitted power at $f_2$ is now defined as the sum of the powers at $f_2$ exiting the device through the channels indexed 1 and 2.

We found this optimization to be more difficult with the available tunable degrees of freedom in our experiment. On the one hand, the result from Fig. 1C suggests that a purely frequency-constrained RSM at $f_1$ may be easier to find with $N_{in} = 2$ than with $N_{in} = 1$. But, on the other hand, the simultaneous satisfaction of additional signal-routing constraints may turn the multi-channel advantage into a disadvantage. Moreover, undesirable transmission toward two channels rather than one channel must now be avoided at $f_2$. To ease the optimization burden, we allow some flexibility regarding the chosen operation frequencies in our optimization protocol. Of course, this is not a general limitation of our concept but related to the specific constraints of our proof-of-concept experiment.

Three selected implementations of programmable reflectionless wavelength multiplexing with multi-channel excitation for one of the two operating frequencies are shown in Fig. 4A. The inset illustrates the corresponding required adapted wavefronts and Table 3 summarizes the performance metrics. We achieve good reflection suppression (at least 48 dB) and transmission discrimination (at least 30 dB). In this multi-channel scenario, it is interesting to explore the sensitivity to phase or amplitude detuning of the adapted wavefront. In Fig. 4B, we measure *in situ* how the reflected, undesirably transmitted, and desirably transmitted powers at $f_1$ are affected by detuning of the phase or amplitude of the adapted wavefront, or detuning of the frequency. Any detuning increases the reflected power, including the (possibly at first glance surprising) case of excitation with lower amplitude on one channel. This confirms that we have indeed implemented a (functionalized) RSM. The undesired transmitted power can in some cases be slightly lower upon small detuning, implying that our optimization imposed a close-to-zero-scattering condition for these undesirable transmissions, evidencing again that in the optimizations underlying Fig. 4 it was harder to fully satisfy all six constraints simultaneously. Overall, in line with our conclusions from Fig. 2, we observe that the reflected and undesired transmitted power are very sensitive to any type of detuning whereas the desired transmitted power is very insensitive. This agrees with our conjecture above that a zero-scattering condition is a very narrow-band spectral singularity, while a maximum-scattering-amplitude condition can be achieved without sensitive interference and hence is less sensitive to detuning.

| $f_1$ [GHz] | $f_2$ [GHz] | $P_R$ [dB] | $P_{Tu}$ [dB] | $P_{Td}$ [dB] | D = $P_{Td}$ - $P_{Tu}$ |
|---|---|---|---|---|---|
| **5.05** | 5.245 | −52 | −56 | −21 | 35 |
| 5.245 | **5.35** | −54 | −51 | −21 | 30 |
| 5.23 | **5.305** | −48 | −55 | −21 | 34 |

**Table 3. Performance metrics of the *in-situ*-observed multiplexer functionalities displayed in Fig. 4.** The frequency injected through two ports is highlighted in bold font.

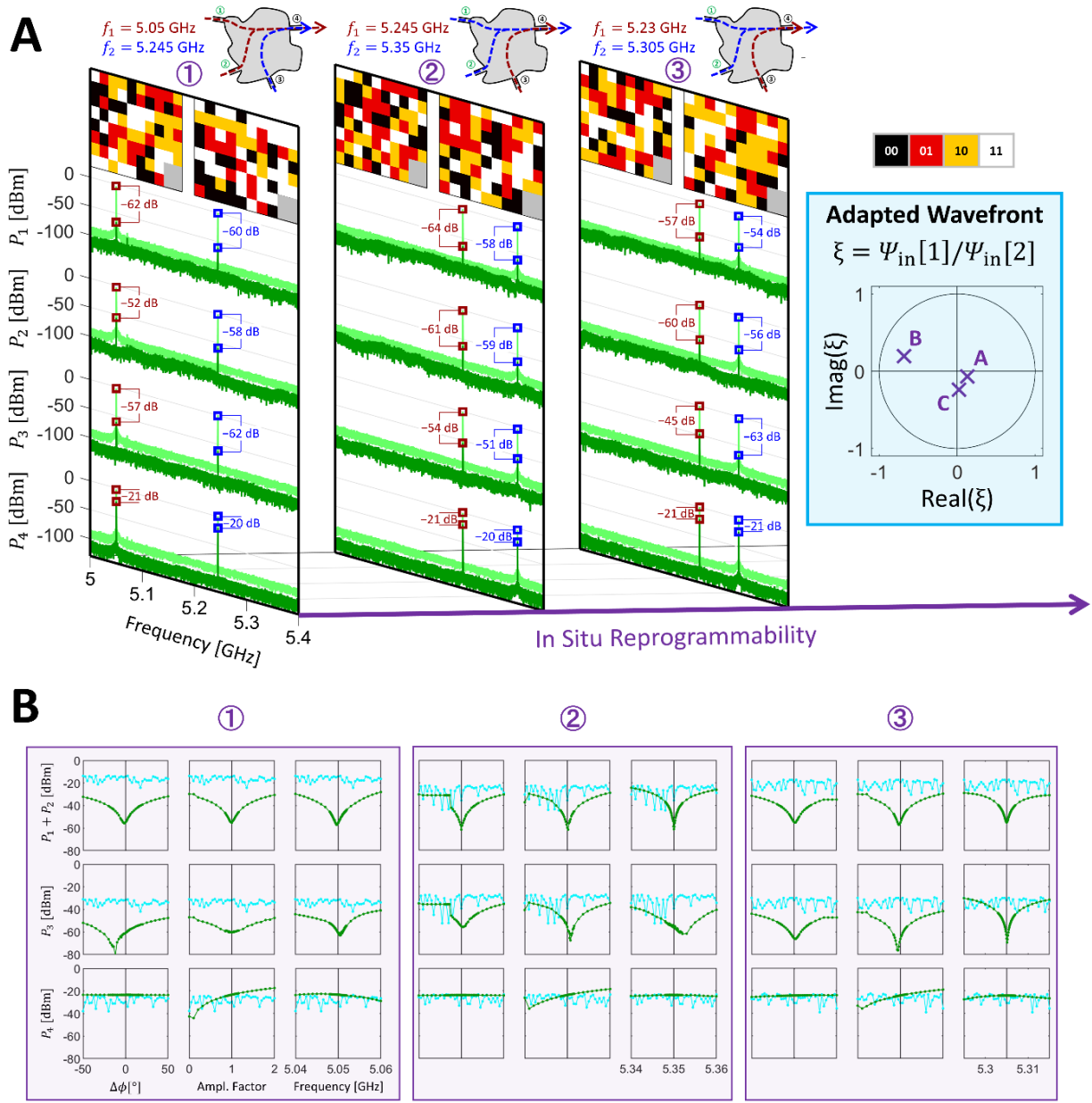

**Fig. 4. Programmable reflectionless wavelength multiplexer with multi-channel excitation requiring an adapted input wavefront.** (**A**) Three instances of *in situ* observations of a reflectionless multiplexer for different pairs of $f_1$ and $f_2$, where one of the two frequencies is injected via two ports, hence requiring an adapted input wavefront. The corresponding two-channel input wavefronts are visualized in the inset in terms of the relative phase and amplitude difference. Table 3 summarizes key performance characteristics. (**B**) Analysis of sensitivity to detuning of the adapted wavefront. For each of the three instances from (A), we observe *in situ* how the reflected ($P_1+P_2$), undesirably transmitted ($P_3$) and desirably transmitted ($P_4$) power varies upon detuning the relative phase (left) or amplitude (middle) of the adapted wavefront. We also check frequency detuning (right). Only results for the

frequency injected through two ports are plotted (in green) in each case. The artifact upon phase detuning that is apparent in the second instance (②) originates from our coherent source's limited ability to generate arbitrary relative phase differences. We also plot (in light blue) the same three quantities as observed *in situ* upon injecting arbitrary random (but normalized) wavefronts; these values are for reference only and the horizontal axes bear no relevance for them.

## Discussion

The need for reflectionless programmable signal routing is ubiquitous in wave engineering. Our proof-of-principle experiment is one specific implementation of our generic concept. The theory is completely general and only requires a complex cavity to serve as the routing region, massively parametrized by suitable external control knobs to explore a large ensemble of scattering matrices. Generalizations to guided or trapped waves in higher frequency regimes appear to be already within reach (*35, 57, 58*). Incidentally, our experimental setup maps directly into that of recently proposed metasurface-programmable wireless networks-on-chip (*43*). But, in principle, the connected channels can be guided or freely propagating in nature. For instance, one or multiple connected channels could be antennas that radiate waves to the far field. Then, based on our concepts, guided waves at different frequencies can be converted to freely propagating waves radiated by specific antennas without any reflection. This use case relates to Ref. (*59*) where a variant of CMR was leveraged to feed antennas with reduced reflection; however, Ref. (*59*) did not establish the connection with well-known coherent wave control concepts nor did it attempt to optimize or parametrize the scattering system itself to achieve reflectionless and/or programmable signal routing. It is also interesting to consider a multi-functional antenna itself as a signal router that converts incoming signals from guided mono-modal waveguides (for radiofrequency antennas such as Ref. (*60*)) or from quantum emitters (for nanoantennas) to beams propagating in free space. Clearly, reflectionless conversion of signals at different frequencies into different beams, in an *in-situ* reprogrammable manner, would be highly desirable. First attempts at improving radiation efficiency by designing the scattering environment (*61*) can be interpreted as aiming at suppressing reflections as the source couples radiation to the environment, but features like routing and programmability are still missing. Such antenna concepts involve both guided and radiative channels, and care must be taken to ensure that the channels form an orthogonal basis. Besides the nature of the input-output channels, the scattering system itself can take diverse forms in our concept. Instead of using a chaotic cavity, our concept could, for instance, also be implemented based on networks of transmission lines such as quantum graphs (*62*) or Mach-Zehnder-interferometer meshes (*63*), provided that a technique to massively parametrize them is available.

The reflectionless aspects of our signal routing concepts are based on a rigorous underlying theory which implies that with one or a few tuning parameters, and wavefront shaping if $N_{in}>1$, reflectionless scattering can be achieved under rather general conditions. The results in Fig. 1 confirm the expectations for RSMs within that theory. The ability to optimize RSMs for additional functionalities or with constraints does not yet have a corresponding conceptual and analytic framework, although the results shown here indicate a rather general capability to satisfy multiple additional constraints while maintaining an RSM. Specifically, it would be helpful to understand how the ability to satisfy constraints improves as a function of the number of independent tuning parameters and how this evolves as the system moves from isolated resonances to the regime of overlapping resonances studied here. These and related questions represent an important direction for future research.

Our results may appear reminiscent of open and closed channels in diffusive multiple-scattering media (*64*, *65*) but the underlying principles of our reflectionless signal routers are more general and independent of the scattering system's geometry. Open and closed channels in diffusive media originate from the bimodal distribution of eigenvalues of the system's *S*-matrix; injection of suitable adapted wavefronts enables transmission of order unity or low transmission, respectively, at any frequency (since the frequency does not even enter the calculation to derive the bimodal eigenvalue distribution). However, such open channels do not achieve arbitrarily small reflection; the latter requires tuning of the scattering system. By tuning or optimizing the scattering system, an *R*-zero can be placed arbitrarily close to the real frequency axis, in line with RSM theory, but this feature then only exists for a discrete frequency as opposed to at any frequency. Moreover, note that RSM theory is only based on the analytic properties of the scattering operator and makes no assumption about the system geometry being diffusive. Indeed, our experimental cavity is not a diffusive system.

The perfect suppression of reflection and undesired transmission is inherently only possible at the discrete frequencies at which the zero-scattering singularities are imposed. Moreover, as seen in Fig. 2 and Fig. 4, the zero-scattering conditions are very sensitive to frequency detuning. Therefore, the bandwidth of a given signal that can be routed with *perfect* suppression of reflection and undesired transmission is fundamentally limited in linear passive time-invariant devices. While this is not necessarily an inconvenience for signal routers used to deliver energy with CW signals, it can be problematic for signal routers used to transfer information with modulated CW signals. Given the high density of zeros and massive parametrization of our concept, we will explore in future work the possibility of placing multiple zeros close to each other on the real frequency axis in order to achieve broadband near-perfect zero scattering.

Looking forward, timely research threads lay out multiple additional ideas to achieve near-reflectionless broadband programmable signal routing. First, exceptional points (EPs) of scattering zeros (as opposed to scattering poles) have a flatter frequency response than simple scattering zeros, i.e., the near-zero behavior extends over a larger band (*29*). Recent experiments (*66*, *67*) have confirmed this theoretical prediction in systems with special geometries or symmetries but how to tune an arbitrary complex scattering system to a zero-scattering EP remains an open challenge. Second, by using nonlinear and/or active and/or time-varying systems, fundamental limitations of linear passive time-invariant systems like ours can be overcome (see Ref. (*68*) for an overview of recent work along these lines). Third, additional opportunities may arise from modulating the input signals. For instance, zeros away from the real frequency axis can be accessed through complex-frequency excitations such that reflections are perfectly suppressed for a finite time (*69–71*). Carefully engineered systems may also be able to simultaneously impose zeros of different filtered scattering matrices at the same complex frequency (*72*). Fourth, there are interesting links to the recently explored use of doped epsilon-near-zero media for (not necessarily perfect) impedance matching where the dispersion of the effective permeability is tuned to tailor the bandwidth (*73*).

To summarize, we demonstrated experimentally that massively parametrized overmoded scattering systems can be tuned to functionalize Reflectionless Scattering Modes such that they enable reflectionless programmable signal routing. We demonstrated with *in-situ* observations that both operating frequencies and routing functionality can be flexibly reprogrammed. From a technological perspective, our agile router's ability to avoid devastating reflected-power echoes in networks should remove the need for currently used additional components (isolators, circulators, in-line attenuators). From a more fundamental perspective, our simulations reveal a large broadening of the distribution of reflectionless scattering frequencies in the complex plane as the system becomes strongly overdamped. Without this effect it is unlikely that functionalization of RSMs would be possible. We attribute this effect to differential absorption rates for different eigenstates based on their spatial distribution, and also to increased eigenvalue interactions. A more quantitative theory for this broadening, based on rigorous scattering calculations and statistical analysis to better understand these phenomena are an important avenue for future theoretical work in mesoscopic wave physics (see also Supplementary Note 4).

## Methods

**Experimental setup.** Our experimental setup is shown in Fig. 1B and is based on an electrically large irregularly shaped metallic box with dimensions 0.385m × 0.422m × 0.405m. 16% of its boundaries are massively parametrized by two programmable metasurfaces, each comprising 152 meta-atoms. Each meta-atom has a 1-bit programmable reflection response that roughly mimics Dirichlet or Neuman boundary conditions for one field polarization (see Fig. S1) (*74*). Each unit cell fuses two such meta-atoms, one rotated by 90°, and is hence 2-bit programmable (1 bit for each polarization), explaining the 2-bit color code used in Fig. 3 and Fig. 4A to visualize the metasurface configurations. Four single-mode guided scattering channels are connected to the system via waveguide-to-coax adapters which are broadband impedance matched in free space (roughly frequency-flat reflection coefficient around −15 dB in free space, see Fig. S2). The scattering system has a composite quality factor of 369, and 23 modes overlap at a given frequency. The average power transmitted between two scattering channels through our scattering system is −28.4 dB. Further details and characterizations of our setup are provided in Supplementary Note 1.

Our measurement apparatus can be switched between two measurement modalities: measuring the system's full scattering matrix with a vector network analyzer or performing *in-situ* observations. The latter consist in injecting CW signals at $f_1$ and $f_2$, coherently controlled for $f_1$ in the case of Fig. 4, and measuring the power exiting our system through all connected channels with a spectrum analyzer. *In-situ* observations are important to provide direct experimental evidence of the claimed routing functionalities, as opposed to relying on the linearity of the wave equation to use simulations based on measured scattering matrices. Further details can be found in Supplementary Note 1, including in Fig. S5 a photographic image of the measurement apparatus underlying Fig. 4.

**Calibration.** To avoid that cables and other component of our measurement apparatus impact the characterization of our scattering system, we perform multiple calibrations. First, the VNA is calibrated using a standard electronic calibration kit to remove the effect of propagation from the VNA ports to the circulators. Second, calibration measurements with the *in-situ* measurement setup are performed to measure the signal attenuation due to the measurement apparatus. Specifically, we sequentially connect with an in-line SMA adapter each cable through which energy is injected directly to each possible cable through which energy exits. The received power on the spectrum analyzer in this case which does *not* involve propagation through the scattering system serves as calibration of the measurement apparatus for *in-situ* observations and the corresponding measurements are plotted in light green in Fig. 2D, Fig. 2F and Fig. 2H, Fig. 3 and Fig. 4A. Further details are provided in Supplementary Note 2.

For the injection of the adapted wavefront in Fig. 4, we additionally perform a third calibration to determine relative amplitude and phase shifts between the two sources due to possible inaccuracies inside the multi-channel coherent signal generator and/or differences in the cables between signal generator and scattering system. The detailed procedure is summarized in Supplementary Note 2.

**Optimization.** The identification of a metasurface configuration that yields a desired scattering response of our massively parametrized system is very challenging because no forward model exists to map a given metasurface configuration to the corresponding scattering response. Given the impossibility of accurate analytical forward models, training an artificial neural network to approximate a forward model may be possible in the future. This was already successfully achieved for weakly scattering perturbations in a multi-mode waveguide (*75*), similar to the setup from Ref. (*35*). However, the parametrization of scattering systems with strong reverberation such as ours is highly non-linear because the impact of a given meta-atom on the scattering response depends on how the other meta-atoms are configured (*76*). Moreover, the 1-bit programmability severely limits gradient descent techniques since the tunable parameters can only be chosen from a discrete set rather than from a continuous range. In this paper, we therefore used a simple iterative trial-and-error algorithm based on VNA measurements of the system's scattering matrix to identify a suitable metasurface configuration. Details on the weights used to trade off the six constraints in the cost function as well as an algorithmic summary can be found in Supplementary Note 3. The identification of suitable metasurface configurations for various routing functionalities and operating frequency pairs can be performed offline in a calibration phase such that this optimization does not thwart the runtime deployment of our reflectionless programmable signal router.

**Numerical study.** The simulated structure in the inset of Fig. 1D is a two-dimensional 3-port rectangular cavity with three scatterers. The cavity dimensions are $L \times \frac{3}{4}L$ in the x-y plane. Similar to Refs. (*28*, *43*), we apply an Impedance Boundary Condition (IBC) at the walls of the cavity, which simulates a metallic wall followed by an infinite domain. The domain has a unit relative permeability and permittivity. We can alter the amount of absorption by the walls of the cavity by changing the conductivity: lower values of conductivity correspond to increasingly absorptive walls. The scatterers are circular and perfectly conducting. For different ratios of absorption coupling to radiation coupling, we use an eigenfrequency solver to find the complex-valued reflection zeros of the scattering system using a perfectly matched layer method within a frequency interval from 50 $c/L$ to 75 $c/L$. Additional details, explanations and results can be found in Supplementary Note 5.

## Supplementary Materials

Supplementary Note 1: Details on the Experimental Setup

Supplementary Note 2: Details on the Calibration of the Measurement Setup

Supplementary Note 3: Details on the Optimization of the Metasurface Configuration

Supplementary Note 4: Additional Analysis of Unconstrained RSMs

Supplementary Note 5: Additional Details on the Numerical Simulations

## Acknowledgments

A.D.S. acknowledges useful discussions with M. Fink and F. Lemoult. P.d.H. acknowledges stimulating discussions with S. Horsley, D. B. Phillips and R. Sapienza. The metasurface prototypes were purchased from Greenerwave.

## Funding

J.S. and P.d.H. acknowledge funding from the European Union through the European Regional

Development Fund (ERDF), and the French region of Brittany and Rennes Metropole through the CPER Project SOPHIE/STIC & Ondes. A.A and A.D.S. acknowledge the Simons Collaboration on Extreme Wave Phenomena.

## Author Contributions

A.D.S. and P.d.H. conceived the project. J.S. and P.d.H. conducted the experiments. A.A. and A.D.S. conducted the numerical simulations. All authors interpreted the results and contributed with thorough discussions. P.d.H. wrote the manuscript.

## Competing Interests

The authors declare no competing interests.

## Data and materials availability

All data needed to evaluate the conclusions in the paper are present in the paper and/or the Supplementary Materials. Additional data related to this paper may be requested from the authors.

# Supplementary Materials for

# Reflectionless Programmable Signal Routers

Jérôme Sol[1], Ali Alhulaymi[2], A. Douglas Stone[2], Philipp del Hougne[3*]

[1] INSA Rennes, CNRS, IETR - UMR 6164, F-35000 Rennes, France

[2] Department of Applied Physics, Yale University, New Haven, Connecticut 06520, USA

[3] Univ Rennes, CNRS, IETR - UMR 6164, F-35000 Rennes, France

* Correspondence to philipp.del-hougne@univ-rennes1.fr.

Table of Contents:

## Supplementary Note 1. Details on the Experimental Setup

In this supplementary note, we provide further details on the experimental setup.

### *Disordered Metallic Box*

A photographic image of the inside of the disordered metallic box equipped with two programmable metasurfaces and coupled via waveguide-to-coax adapters to three or four single-mode guided channels is shown in Fig. 1B. The metallic box has dimensions of $0.385\mathrm{m} \times 0.422\mathrm{m} \times 0.405\mathrm{m}$ and hence a volume of $0.0658\mathrm{m}^3$.

Based on the decay rate of the inverse Fourier transform of transmission spectra measured between different ports for a series of random metasurface configurations, we estimate the system's composite quality factor as $Q = 369$. Our estimate of $Q$ does not significantly vary if we disconnect one or multiple channels from the system, implying that the decay rate is strongly dominated by absorption rather than out-coupling. This observation is in line with the fact that the transmission coefficient has an average magnitude of $-28.4$ dB.

Based on Weyl's law, we find that around $\mathcal{N} \sim \frac{8\pi V}{c^3 Q} f_0^3 = \frac{8\pi}{Q} \frac{V}{\lambda_0^3} = 23$ modes overlap at a given frequency within the considered interval. We operate thus in a multi-resonance transport regime that cannot be understood by applying intuition derived from systems with a single isolated resonance.

The irregularity of the shape of the metallic box is achieved through two mechanisms. On the one hand, as seen in Fig. 1B, metallic scattering structures (a cylinder piercing into the volume of the cavity and an eighth of a sphere in one of the cavity's corners) perturb the regular box geometry and breaks its symmetries. Note that also the metallic waveguide-to-coax adapters have such a geometry-perturbing effect. On the other hand, we impose irregular coding patterns on the programmable metasurface which can be thought of as an electronic equivalent of irregular cavity boundaries (*1*), adding further to the shape irregularity.

### *Programmable Metasurface*

Programmable metasurfaces are ultrathin arrays of meta-atom with electronically reconfigurable scattering properties. The concept can be traced back to early works in the 2000s (*2*, *3*) and received renewed attention since 2014 (*4*, *5*). Programmable metasurfaces are also referred to as "tunable impedance surfaces", "spatial microwave modulators", or "reconfigurable intelligent surfaces".

In our present work, we use a programmable metasurface whose constituent meta-atoms have a 1-bit programmable reflection coefficient. In other words, each meta-atom can be individually toggled via a DC bias voltage between two possible states ("0" and "1") whose reflection coefficients roughly mimic Neuman or Dirichlet boundary conditions. Our metasurface prototypes (purchased from Greenerwave) are based on the design introduced in Ref. (*5*). Therein, each meta-atom consists of two resonators that hybridize, and the bias voltage of the PIN diode controls the resonance frequency of one of these two resonators. Thanks to this mechanism, the phase of the reflected wave can be altered by roughly $\pi$ for one field polarization. In our prototype, two such meta-atoms, one rotated by 90°, are fused such that each unit cell has independent control over both field polarizations. Therefore, each unit cell is 2-bit programmable (1 bit for each polarization), and consequently our color code in Fig. 3 and Fig. 4 to visualize the metasurface coding patterns contains four possible colors for "00" (black), "01" (red), "10" (yellow) and "11" (white).

It is important to note that our present work does not rely on the specific utilized metasurface design. The key hardware ingredient of our work is a *massively parametrized* overmoded complex scattering system. The massive parametrization is conveniently achieved with a programmable metasurface in our experiments. But the exact nature of the parametrization does not matter, and many other metasurface designs or alternative mechanisms to parametrize a scattering system could be used. In our case, an ideal programmable metasurface (i) interacts with as many rays as possible, and (ii) its programmability is as fine-grained as possible (but at least 1-bit). Interaction with as many rays as possible can be achieved by using meta-atoms with the largest possible scattering cross-section and using as many meta-atoms as possible.

To characterize the utilized programmable metasurface prototype (*6*), the reflection coefficient of the metasurface is measured with waves that are normally incident and polarized along one of the two meta-atoms. Our measurements are conducted with the horn-antenna setup shown in Fig. S1A, once with all meta-atoms in state "1" and once with all meta-atoms in state "0". The experimentally measured frequency-dependent magnitude and phase responses are plotted in Fig. S1B and Fig. S1C, respectively. Around 5.15 GHz, the phase response shows a roughly $\pi$ phase difference between the two possible states, while the magnitude response is roughly the same. Note that the absolute value of the magnitude response depends on the utilized horn antenna such that this data cannot be used to quantify the amount of energy that is absorbed by the metasurface.

Because waves from all possible angles impinge upon our metasurface in our experiment, we further report in Fig S1D the results from an *in-situ* characterization of the metasurface prototype's ability to manipulate the complex field inside our scattering system. With both programmable metasurfaces placed at their intended locations, the reflection coefficient at one port for 500 random metasurface configurations is measured and the standard deviation across these 500 complex-valued measurements is plotted. Now we can clearly see that the programmable metasurface efficiently modulates the field in the vicinity of 5.15 GHz within an interval of roughly 400 MHz.

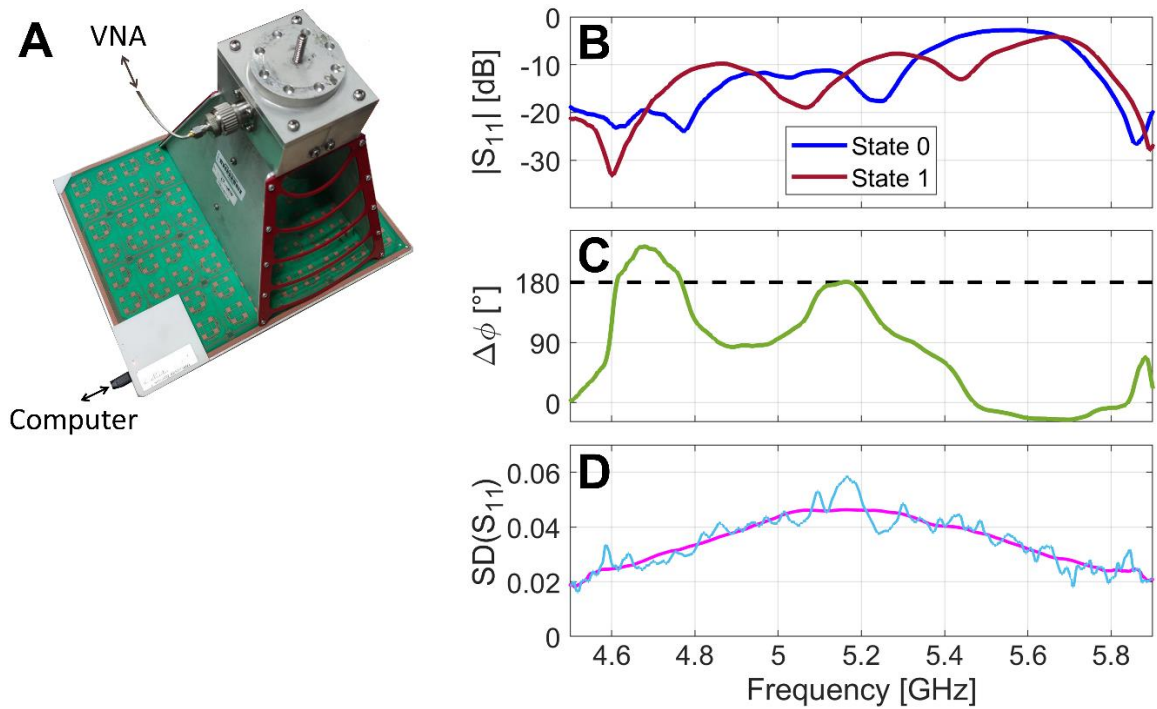

**Fig. S1. Characterization of the programmable metasurface.** (**A**) Measurement setup to characterize the metasurface's response under normal incidence for one polarization. (**B**) Response magnitude measured with the setup from (A) when all meta-atoms are simultaneously either in state "0" or state "1". (**C**) Phase difference between the responses measured with the setup in (A) in the two possible states. (**D**) *In-situ* characterization of the metasurface's ability to manipulate the complex electromagnetic field via the standard deviation of the reflection spectrum measured inside the disordered metallic box for 500 random metasurface configurations. A smoothed version of the curve is also shown.

## *Waveguide-to-Coax Adapters*

The reflection coefficient of the waveguide-to-coax adapters (RA13PBZ012-B-SMA-F) used in our experiment to couple the scattering channels to the scattering system (see Fig. 1B) is almost frequency-flat in an anechoic environment at a value around $-14.8$ dB, as seen in Fig. S2.

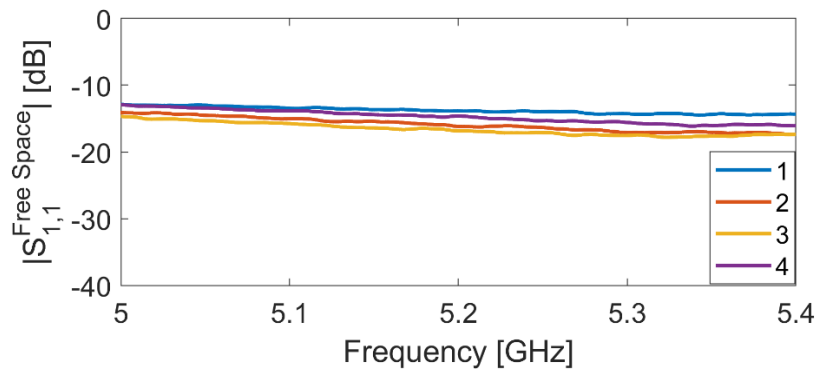

**Fig. S2.** Reflection coefficient of the four waveguide-to-coax adapters measured in an anechoic environment.

## *Separation of Injected and Exiting Signals*

For our *in-situ* observations, it is essential to separate the signals exiting the scattering system toward the spectrum analyzer from the injected signals coming from the signal generators. We achieve this separation by connecting each coax-to-waveguide adapter through which signals are injected to a circulator (PE83CR006). Circulators are standard off-the-shelf components in microwave engineering. Details of how the circulators are integrated into our three experimental setups can be seen in the respective schematic drawings in Fig. 2B, Fig. S3 and Fig. S4, as well as in the photographic image in Fig. S5 corresponding to the schematic from Fig. S4.

## *Switching between VNA and In-Situ Measurements*

Our ultimate goal are *in-situ* observations of functional RSMs that enable signal routing.

Page 5 of 25

But for the auxiliary steps that are necessary to identify a suitable metasurface configuration for a desired pair of operating frequencies and signal routing functionality, it is more convenient to measure the system's scattering matrix with a VNA. The main reason is that the VNA measures the entire scattering matrix much faster. To accommodate both measurement modalities (VNA and *in-situ* observation), our measurement setup is capable of being switched electronically between connecting the scattering system to a VNA or to the *in-situ* measurement apparatus (involving signal generators and spectrum analyzer). This switching is achieved with up to seven 1-to-2 RF switches (PE71S6436) and can hence be orchestrated by the host computer. A major advantage over manual switching is that accidental perturbations of the measurement setup are avoided. Details of how the RF switches are integrated into our three experimental setups can be seen in the respective schematic drawings in Fig. 2B, Fig. S3 and Fig. S4, as well as in the photographic image in Fig. S5 corresponding to the schematic from Fig. S4.

Due to the presence of circulators to enable the *in-situ* observations, the VNA effectively faces a system with four, five or seven ports for our three experimental setups (see schematic drawings in Fig. 2B, Fig. S3 and Fig. S4) even though they only involve three or four channels coupled to our scattering system. For instance, a simple single-channel reflection measurement is converted into a transmission measurement through a circulator. The VNA thus measures a $4 \times 4$, $5 \times 5$ or $7 \times 7$ scattering matrix and we subsequently select the relevant entries to constitute the sought-after $3 \times 3$ or $4 \times 4$ scattering matrix.

*Measurement Setup Configurations*

We have worked with three distinct experimental setups to obtain the results from
  (i)   3-port wavelength demultiplexer (Fig. 2 and Fig. 3A).
  (ii)  3-port alternative signal routing functionality (Fig. 3B).
  (iii) 4-port wavelength multiplexer with one frequency being injected through two channels (Fig. 4).

A schematic of the experimental setup underlying (i) is provided in Fig. 2B. The schematics for (ii) and (iii) are provided in Fig. S3 and Fig. S4, respectively. In addition, a photographic image of the experimental measurement setup corresponding to (iii) is provided in Fig. S5.

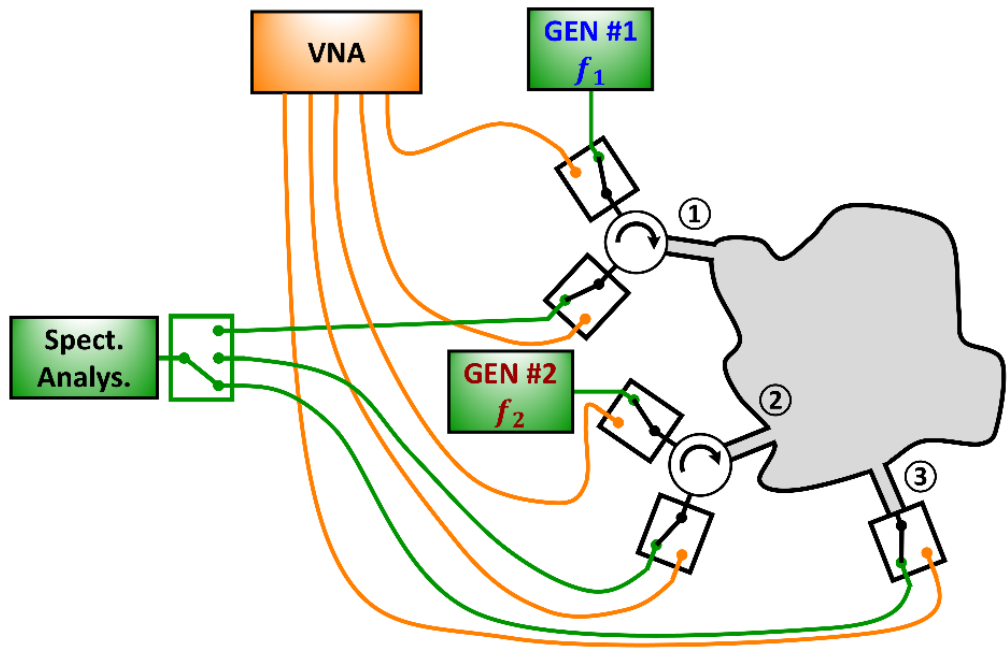

**Fig. S3.** Schematic drawing of the experimental setup underlying Fig. 3B involving both VNA measurements and *in-situ* observation.

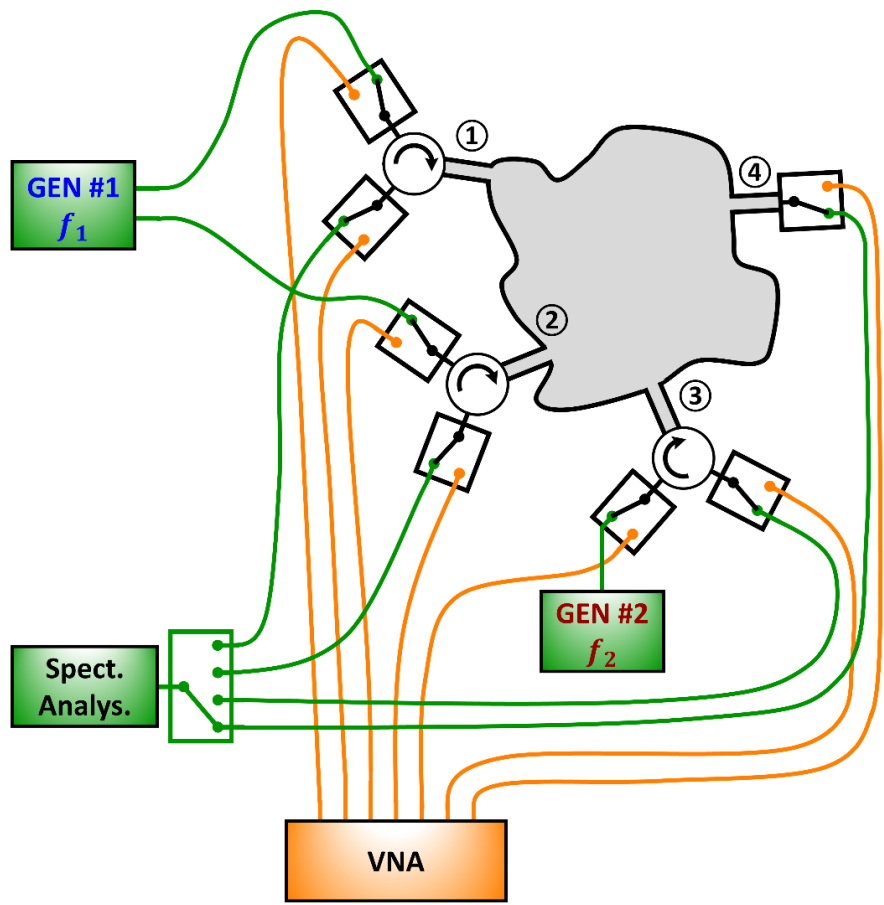

**Fig. S4.** Schematic drawing of the experimental setup underlying Fig. 4 involving both VNA

measurements and *in-situ* observation. Note that GEN #1 coherently controls the multi-channel output of $f_1$.

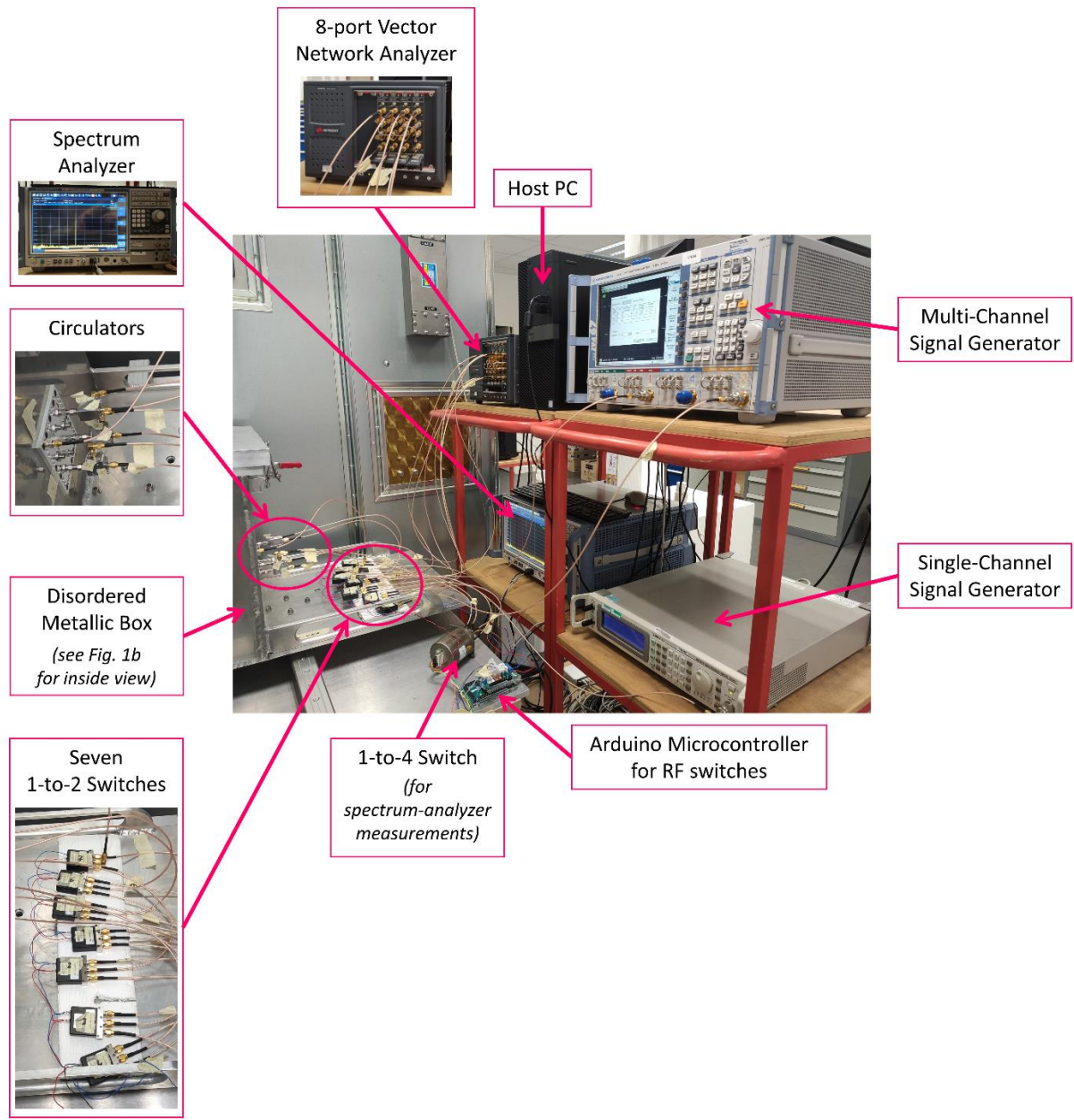

**Fig. S5.** Photographic image of the measurement setup corresponding to the schematic drawing from Fig. S4 of the experimental setup underlying Fig. 4. The measurement setup includes an 8-port vector network analyzer (Keysight M9005A chassis with four M9374A modules, 300 kHz – 20 GHz), a multi-channel signal generator (Rhode & Schwarz ZVA 67, DCM, 10 MHz – 67 GHz) capable of generating an adapted wavefront, a single-channel signal generator (Aeroflex IFR 3416, 250 kHz – 6 GHz), a spectrum analyzer (Rhode & Schwarz FSW, 2 Hz – 26.5 GHz), three circulators (PE83CR006, 4 GHz – 8 GHz) to separate injected and reflected signals, seven 1-to-2 switches (PE71S6436, DC – 18 GHz) to switch between measuring the scattering matrix with the 8-port VNA and performing the *in-situ* measurements with signal generators and spectrum analyzer, a 1-to-4 switch (Dow-Key Microwave 581 J-520803A, DC – 18 GHz) to measure the scattering system's output signals successively on the spectrum analyzer, an Arduino microcontroller to configure all the switches, and a host computer orchestrating the entire measurement procedure.

**Supplementary Note 2**. Details on the Calibration of the Measurement Setup

In this supplementary note, we discuss the details of how our measurement setup is calibrated. Our experiments involve two important calibrations. The first calibration concerns absorption in the cables of the measurement setups (relevant to all *in-situ* observations). The second calibration concerns the injection of an adapted wavefront (only relevant to the *in-situ* observations in Fig. 4).

*Signal Absorption in Cables of the Measurement Setup*
Our goal is to measure *in situ* the signal power exiting the scattering system through the various connected channels. However, the cables leading from the signal generators to the scattering system as well as from the scattering system to the spectrum analyzer introduce additional signal attenuation that is not due to the scattering system and hence not of interest. Therefore, we measure the signal attenuation in the cables of our measurement setup in a calibration step. Specifically, for each pair of input channel (only those connected to a signal generator) and output channel, we disconnect the two cables from the scattering system and directly connect them with an in-series SMA adapter. We then emit the same unity CW signal as during the main experiment, and we observe the received power on the spectrum analyzer. This procedure is repeated for each considered frequency.

Because the transfer function of a cascade of linear systems is the product of the transfer functions of each individual system, we can deduce the transfer function of our scattering system of interest by dividing the transfer function of the measurement setup connected to the scattering system by the transfer function of the measurement setup without connection to the scattering system. On the logarithmic dBm scale, this division straightforwardly corresponds to a simple subtraction, as seen in Fig. 2, Fig. 3 and Fig. 4. Our *in-situ* observation is hence independent of what signal levels are defined as unity at the signal generators.

For the case of the adapted wavefront in Fig. 4, we let the two-channel source emit a unity CW signal on each of the two channels in turn, and we measure the corresponding outputs as before. However, because this corresponds to injecting 2 a.u. instead of 1 a.u. of power at the considered frequency, we divide the measured outputs by 2. Note that the power injected at the considered frequency in the main experiment is always 1 a.u. because the RSM wavefront is normalized.

*Injection of an Adapted Wavefront*

Our goal in Fig. 4 is to inject a two-channel adapted wavefront into our scattering system, meaning that upon injection there should be a specific relative phase and amplitude difference $\xi = \Psi_{in}[1]/\Psi_{in}[2]$ between the two entries of $\Psi_{in}$. In practice, inaccuracies inside the multi-channel coherent signal generator and/or differences in cables between the signal generator and the scattering system can alter $\xi$. To avoid the latter, we perform an additional calibration step.

To determine the relative phase and amplitude errors due to sources and cables, we disconnect the two cables that inject the two-channel wavefront into our scattering system from the latter, and instead we connect them via a power combiner (PS2-20-450/10S) directly to the spectrum analyzer. First, we emit wavefronts $\Psi_{10} = [1 \quad 0]$ a.u. and $\Psi_{01} = [0 \quad 1]$ a.u., and we measure the corresponding powers $P_{10}$ and $P_{01}$. The multiplicative amplitude correction factor for the second channel is then simply $\zeta_A = \sqrt{P_{10}/P_{01}}$ (with powers in linear units). We find values of $\zeta_A$ very close to unity in all cases. Second, with the amplitude correction implemented, we emit wavefronts $[1 \quad \zeta_A e^{i\phi}]$ a.u. for 20 linearly spaced values of $\phi$ between $-162°$ and $162°$, and we measure the corresponding powers $P(\phi)$. The reason for using $\pm 162°$ instead of $\pm 180°$ relates to limitations of the coherent multi-channel signal generator to impose arbitrary relative phase differences. We then identify the additive phase correction term for the second channel as the value of $\zeta_P$ for which $2\left|\cos\left(\frac{\phi+\zeta_P}{2}\right)\right|$ most closely resembles the measured $\sqrt{P(\phi)}$ curve in terms of the Pearson correlation coefficient.

In the experiments underlying Fig. 4, we multiply the second entry of the input wavefront by $\zeta_A e^{-i\zeta_P}$ to correct for the relative phase and amplitude differences due to sources and cables.

# Supplementary Note 3. Details on the Optimization of the Metasurface Configuration

In this supplementary note, we discuss procedural details of the identification of suitable metasurface configurations for the desired reflectionless signal routing functionalities. This involves both the definition of a suitable cost function as well as an algorithm to minimize that cost function.

*Cost Function*

The desired signal-routing functionalities involve six constraints that are ideally simultaneously satisfied:

(1) Minimization of the power $R(f_1)$ that is reflected back to the signal generator(s).
(2) Minimization of the power $T_\mathrm{u}(f_1)$ that is transmitted to the undesired channel(s).
(3) Maximization of the power $T_\mathrm{d}(f_1)$ that is transmitted to the desired channel(s).
(4) Minimization of the power $R(f_2)$ that is reflected back to the signal generator(s).
(5) Minimization of the power $T_\mathrm{u}(f_2)$ that is transmitted to the undesired channel(s).
(6) Maximization of the power $T_\mathrm{d}(f_2)$ that is transmitted to the desired channel(s).

Our cost function must hence find a trade-off between these six objectives. We found heuristically that $\mathcal{C} = \max(\mathcal{C}_1, \mathcal{C}_2)$ works well, where

$$\mathcal{C}_1 = \max\left( \sqrt{R(f_1)}, \sqrt{T_\mathrm{u}(f_1)}, 0.3 \max(0.1 - \sqrt{T_\mathrm{d}(f_1)}, 0) \right)$$

$$\mathcal{C}_2 = \max\left( \sqrt{R(f_2)}, \sqrt{T_\mathrm{u}(f_2)}, 0.3 \max(0.1 - \sqrt{T_\mathrm{d}(f_2)}, 0) \right).$$

Note that we have not heavily optimized this definition of the cost function.

*Optimization Algorithm*

The parametrization of the scattering matrix through the metasurface configuration is highly nonlinear (7, 8) and not analytically known due to the complex geometry. In other words, we cannot predict the scattering matrix that is obtained with a particular metasurface coding pattern. We also cannot measure the scattering matrix corresponding to all $2^{304}$ possible metasurface configurations in order to subsequently identify the globally optimal one with respect to our cost function. Moreover, the 1-bit programmability constraint thwarts conventional gradient-descent optimizations (each parameter can only take one of two possible values rather than being continuously adjustable).

We therefore use an experimental trial-and-error type algorithm to efficiently search the huge parameter space for a satisfactory local optimum. An algorithmic summary is provided below. Therein, the measurements of the scattering matrices are conveniently performed with the 8-port VNA using an emitted power of 0 dBm and an IF bandwidth of 10 kHz. The idea behind this simple optimization algorithm is to begin by determining the cost function for 250 random configurations, in order to pick the best out of these 250 random configurations as starting point for an iterative optimization. Therein, during each iteration the state of one meta-atom is flipped, and the change in configuration is kept if it improves the cost function. We must loop multiple times over each meta-atom because reverberation induces long-range mesoscopic correlations that make the scattering matrix parametrization highly non-linear. In other words, the optimal configuration of a given meta-atom depends on how the other meta-atoms are configured. Like in most inverse design problems where an exhaustive search is unfeasible, there is no guarantee that the identified local optimum is globally optimal. However, we observe that running the algorithm multiple times (with different random initializations) yields local optima of comparable quality. Note that we have not heavily optimized this optimization algorithm.

---
**Algorithm 1:** Binary-Constrained Optimization of $N$-Element Metasurface Configuration
---
1 **for** $i = 1, 2, \ldots, 250$ **do**
2      Define a random binary metasurface configuration $C_i$.
3      Measure the corresponding scattering matrix $S_i$.
4      Evaluate the corresponding cost function $\mathcal{C}_i$.
5 **end**
6 Define $C_\text{curr}$ as the metasurface configuration corresponding to $\mathcal{C}_\text{curr} = \min_i(\{\mathcal{C}_i\})$.
7 **for** $i = 1, 2, \ldots, 5N$ **do**
8      Define $C_\text{temp}$ as $C_\text{curr}$ but with configuration of $\mathrm{mod}(i, N)$th meta-atom flipped.
9      Measure the corresponding scattering matrix $S_\text{temp}$.
10     Evaluate the corresponding cost function $\mathcal{C}_\text{temp}$.
11     **if** $\mathcal{C}_\text{temp} < \mathcal{C}_\text{curr}$ **then**
12         Redefine $C_\text{curr}$ as $C_\text{temp}$ and $\mathcal{C}_\text{curr}$ as $\mathcal{C}_\text{temp}$.
13     **end**
14 **end**
    **Output:** Optimized metasurface configuration $C_\text{curr}$.
---

## Supplementary Note 4. Additional Analysis of Unconstrained RSMs

In this supplementary note, we provide further analysis of the unconstrained RSMs observed in Fig. 1C.

First, we analyze how balanced the RSM wavefronts are in terms of the power injected through the various channels. To this end, we compute two metrics: the entropy, $H$, and participation number, $PN$, of the RSM wavefronts, which are defined as follows:

$$H = \exp\left(-\sum_{i=1}^{N_{\text{in}}} \tilde{\sigma}_i \ln(\tilde{\sigma}_i)\right)$$

$$PN = \frac{\left(\sum_{i=1}^{N_{\text{in}}} |\Psi_{\text{in},i}|\right)^2}{\sum_{i=1}^{N_{\text{in}}} |\Psi_{\text{in},i}|^2}$$

where $\tilde{\sigma}_i = |\Psi_{\text{in},i}| / \sum_{i=1}^{N_{\text{in}}} |\Psi_{\text{in},i}|$, and $\Psi_{\text{in},i}$ is the $i$th entry of the RSM wavefront $\Psi_{\text{in}}$. Both quantities have the same fundamental bounds: the lower bound is unity (corresponds to a wavefront with only one non-zero entry) and the upper bound is $N_{\text{in}}$ (corresponds to a perfectly balanced wavefront). In Fig. S6, we plot the average and standard deviation of these metrics over all unconstrained RSMs identified in Fig. 1C. A clear linear dependence on $N_{\text{in}}$ is evident. We fitted a linear curve to the data and found that its gradient and offset are roughly $\frac{2}{3}$ and $\frac{1}{3}$ for the entropy and roughly $\frac{1}{2}$ and $\frac{1}{2}$ for the participation number.

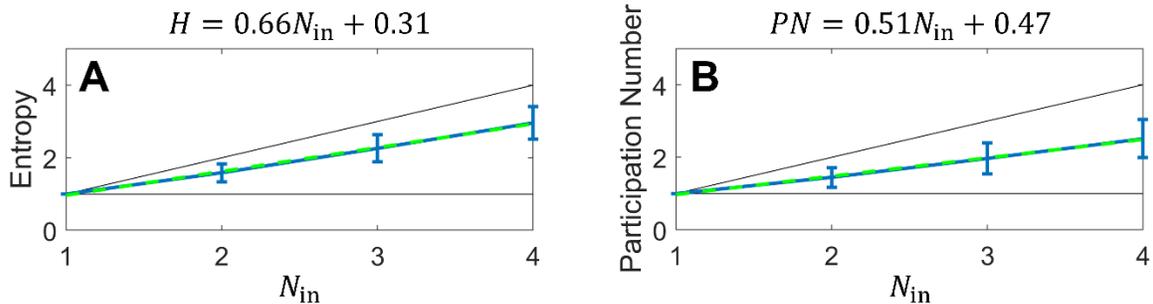

**Fig. S6.** Analysis of the relative weights of the coefficients of RSM wavefronts identified in the analysis underlying Fig. 1C, in terms of their entropy (**A**) and participation number (**B**). $N_{in}$ denotes the number of input channels for the considered RSM. For reference, the lowest (1) and highest ($N_{in}$) values that these two metrics can take are indicated through black lines. Average and standard deviation of these metrics across all identified RSMs for a given value of $N_{in}$ are shown in blue. Additionally, linear fits and their parameters are indicated.

Second, we analyze the probability with which we observe unconstrained RSMs at a given frequency. This analysis includes all data from Fig. 1C (confusing all choices of $N_{\mathrm{in}}$) and the resulting histogram is displayed in Fig. S7. It is apparent that RSMs are much more common at some frequencies than at others. These are characteristics that are specific to our experimental setup, such as the metasurface design, the geometry of the irregular metallic box, and the positioning of the antennas therein.

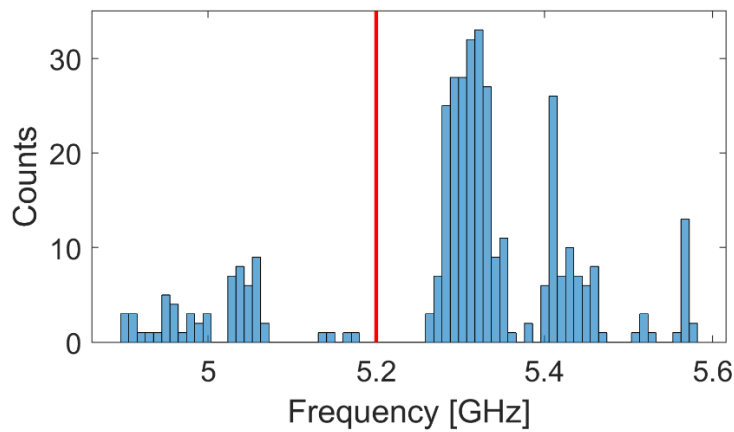

**Fig. S7.** Histogram of the number of RSMs identified at each frequency bin in the analysis underlying Fig. 1C. The vertical red line indicates the targeted frequency of 5.2 GHz discussed in Fig. 1E, Fig. 1F and Fig. 1G.

## Supplementary Note 5. Additional Details on the Numerical Simulations

In this supplementary note, we provide further results and details of our numerical simulations. Unlike our experiments, the numerical simulations provide direct access to complex valued R-zeros and poles. This access to the imaginary part of the eigenfrequencies allows us to perform a statistical analysis on their distributions.

This section is organized as follows. First, we provide additional details on our simulation setup. Second, we summarize the key findings from our simulations. Third, we explore two possible hypotheses to explain the observed eigenvalue spreading in the presence of strong absorption. Fourth, we tune a strongly overdamped system such that it has an RSM using a single continuously tunable parameter. Finally, we demonstrate a functionalized RSM in a unitary scattering system to complement our experimental demonstrations in a scattering system with strong absorption that are reported in the main text.

### *Simulation Setup*

The simulated structure in the inset of Fig. 1D is a two-dimensional 3-port rectangular cavity with three scatterers. The cavity dimensions are $L \times \frac{3}{4}L$ in the x-y plane. Similar to Refs. (*1*, *9*), we apply an Impedance Boundary Condition (IBC) at the walls of the cavity, which simulates a metallic wall followed by an infinite domain. The domain has a unit relative permeability and permittivity. We can alter the amount of absorption by the walls of the cavity by changing the conductivity: lower values of conductivity correspond to increasingly absorptive walls. The scatterers are circular and perfectly conducting, with a radius of $\frac{3}{40}L$. Each of the three leads has a length of $L/2$ and a width of $L/40$. With the center of the cavity defined to be at the origin of our coordinate system, the points at which the centers of the leads meet the cavity are randomly chosen to be $(-L/2, \ 0.2282\ L)$, $(L/2, 0.2966\ L), (L/2, -0.2736\ L)$ and the scatterers positions are arbitrarily assigned as $\left(\frac{3}{16}L, \frac{1}{20}L\right), \left(-\frac{3}{8}L, -\frac{1}{4}L\right)$ and $\left(\frac{1}{80}L, \frac{7}{40}L\right)$.

A perfectly matched layer (PML) of length $L/2$ and width $L/40$ follows each lead. We use an eigenfrequency solver available in a commercial software (COMSOL Multiphysics) to find the complex-valued R-zeros using a modification of the PML method (*10*). Conventional absorbing PMLs are typically used to find resonances in open cavities; their purpose is to suppress reflection at the boundaries of the cavity using a fictious complex scaling function. These absorbing PMLs correspond to outgoing boundary conditions. Purely incoming solutions, on the other hand, can be thought of as the time reverse of purely outgoing solutions, and can be implemented by complex-conjugating the complex scaling function of an absorbing PML. Using a combination of emitting and absorbing PMLs, we can find the different kinds of R-zeros with different boundary conditions. In our case, the software is used to look for angular eigenfrequencies in approximately the rectangular region $(50\ c/L, 75\ c/L) \times (-3.6887\ c/L, 3.6887\ c/L)$, where $c$ is the speed of light in the cavity. The conductivity values explored to obtain the statistics seen in Fig. 1D (except for the lossless case) range between $20\ \epsilon_0 \omega_c$ to $250\ \epsilon_0 \omega_c$, where $\omega_c$ is the central frequency in the search region.

### *Key Findings*

Figure 1D shows that absorption brings down the mean of the imaginary part of the R-zeros. Even though the simple critical-coupling picture cannot be applied in the overdamped regime of this cavity, the imaginary part of the R-zeros can still be interpreted as measuring the degree of undercoupling (upper half frequency plane) or overcoupling (lower half frequency plane). When the R-zero is real (RSM), then by definition that eigenstate (adapted wavefront), at that frequency, is perfectly impedance matched, even though multiple resonances contribute to the in-coupling and out-coupling, and there is no simple interpretation of the impedance-matching in terms of critical coupling. As noted in the main text, an R-zero on the real axis is equivalent to having a zero eigenvalue for the appropriate generalized reflection matrix, determined by the choice of input channels, at real frequency. Hence, the spread of the imaginary part of the R-zeros is the critical metric to study: if the distribution of the imaginary part of the R-zeros (significantly) overlaps zero (i.e., the real frequency axis), then with tuning it is (easily) possible to achieve a system configuration that allows for reflectionless excitation.

We observe that the spread (quantified by the standard deviation in Fig. 1D) widens as the ratio of absorptive coupling to radiative coupling increases. It is this increase in the spread that helps explain how RSMs are found in a regime where absorption loss greatly dominates over radiation loss. In order to quantify the ratio of absorptive coupling to radiative coupling, the radiative coupling is calculated based on the lossless cavity as $\Gamma_{rad} = \langle \text{Im}\{\omega_0 \, L/c\} \rangle$ and the absorptive coupling for each value of the conductivity is calculated as $\Gamma_{abs} = \langle \text{Im}\{\omega_{abs} \, L/c\} \rangle - \Gamma_{rad}$, where $\omega_{abs}$ is the complex zero of the full scattering matrix in the presence of absorption. The average $Q$-factor of the cavity goes down from about 1801 in the lossless case to 38.31 in the highly absorptive case. This 47-fold drop in the $Q$-factor is a measure of the increasingly strong overlap between the resonances in this eventually highly absorptive cavity. As we decrease the conductivity at the walls of the cavity, the absorption coupling goes from zero to $12.5 \, \Gamma_{rad}$, so at the end we are overcoupled by an order of magnitude on average.

While the distribution of the imaginary part of the R-zeros widens at both edges with decreasing conductivity, we find that this broadening is not symmetric; the tail of the distribution is longer at the very negative values than it is nearer to the real axis (see Fig. S8). These outlying eigenvalues push the mean of the distribution further down in the imaginary plane and contribute to the observed high mean absorptive coupling, which then may overestimate the degree of overcoupling, making a fluctuation with tuning or optimization to a zero imaginary part easier to generate.

Before closing this section, we emphasize again that the standard deviations of the distributions of the imaginary part of the R-zeros shown in Fig. 1D are obtained with a random cavity configuration; in contrast, in our experiments we perform a (frequency-constrained) optimization that searches for R-zeros with the desired functional properties in a huge parameter space of possible cavity configurations, and hence possible cavity $S$-matrices. Such a search can come across cavity configurations with anomalous tails of the distribution that are much longer than the one of the randomly chosen cavity configuration we analyzed to produce Fig. 1D. Therefore, our statistical argument explains the plausibility of finding RSMs in strongly overdamped cavities (even without optimization the tails broaden substantially) but does not constitute a direct argument about the ease or difficulty of finding RSMs and functionalizing them.

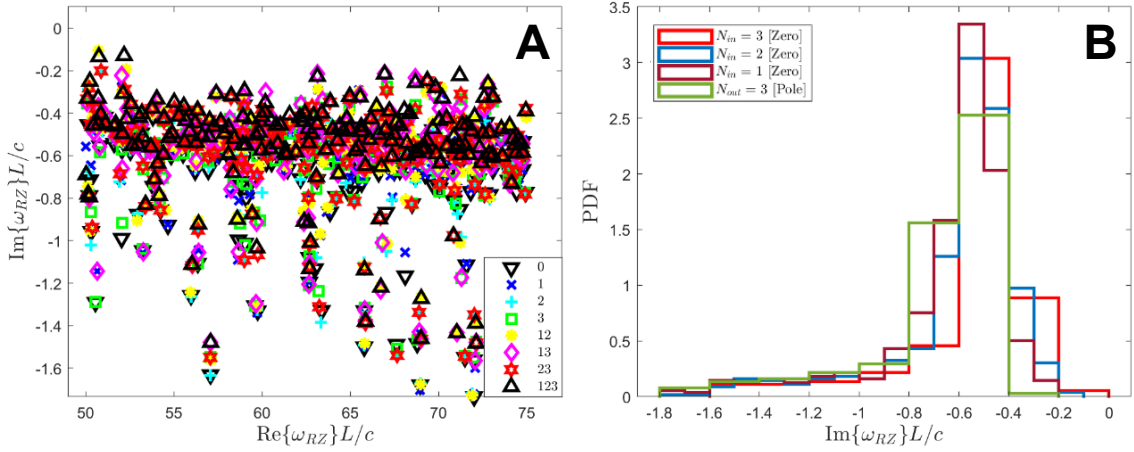

**Fig. S8.** (A) Eigenfrequencies of a system with $\sigma = 36.7\,\epsilon_0\omega_c$. The numbers on the legend correspond to the input ports (e.g., 12 is for the boundary condition with port 1 and 2 taken as inputs), with 0 identifying the pole of the $S$-matrix. (B) A histogram plot of the probability distribution function for the same system as in (A), where we combine the statistics of different choices of input ports with the same number of inputs.

### *Explanations for the Eigenvalue Spreading in Presence of Strong Absorption*

In this section, we present a preliminary investigation of a mechanism that contributes to the spreading of the imaginary parts of the zeros in the strong-absorption regime that we witnessed in Fig. 1D. To this end, we track in Fig. S9A the movement of six zeros in the complex plane with real parts in the range $(55.6,\ 56.7)L/c$. The tracking was performed by sweeping the conductivity nonlinearly in the interval $(50, 50000)\epsilon_0\omega_c$, where $\omega_c$ is the central frequency of the interval over which we track the zeros. One feature of the zeros is that not only do they move downwards due to increased absorption (decreased conductivity), but they also move to lower real parts of the frequency in the complex plane. We attribute this "sideways" motion to the fact that allowing penetration into the walls as conductivity is decreased will effectively increase the wavelength inside the cavity.

*Differential Absorption.* – Introducing finite conductivity into the walls of the cavity should intuitively have a stronger effect on eigenfrequencies with a field that is strongly localized along the walls of the cavity. The imaginary part of the eigenfrequencies will hence depend on $\sim \int_{\text{walls}}|\psi(\mathbf{r})|^2\,d\mathbf{r}$, where $\psi(\mathbf{r})$ is the zero's eigenfield. The field strength at the walls therefore determines the absorption coupling which is expected to significantly impact the speed at which a given zero moves downward in the complex plane. We quantify the overlap of the zero's eigenfield with the absorbing walls via the quantity

$$\mathcal{O} = \frac{\int_{\text{walls}} |\psi(\mathbf{r})|^2 \, d\mathbf{r}}{\int_{\text{cavity}} |\psi(\mathbf{r})|^2 \, d\mathbf{r}} L,$$ where $L$ is the length of the cavity and the integral in the numerator is a line integral along the cavity walls. We observe that $\mathcal{O}$ indeed strongly correlates with the vertical distance covered by the zeros, as seen in Fig. S9B. Indeed, the zero which cover the greatest distance along the imaginary axis has the greatest value of $\mathcal{O}$, and the rest of the zeros are all in the left bottom corner of the plot.

To further support our claim, we plot the inverse participation ratio (IPR) vs. the imaginary part of the zeros of the scattering matrix in Fig. S10. With the IPR defined as $$\frac{\int_{\text{cavity}} |\psi(\mathbf{r})|^4 \, d\mathbf{r}}{\left(\int_{\text{cavity}} |\psi(\mathbf{r})|^2 \, d\mathbf{r}\right)^2},$$ it is a measure of the localization of the field in the cavity. We see from the plot that there is a tendency for zeros with the greatest magnitude of the imaginary part to have a high IPR. A manual inspection of the fields shows that these are mainly localized along the walls of the cavity. Such partially localized states are well-known even in fully chaotic cavities, due to scarring or other wave localization effects, which occur when the system is not deeply in the semiclassical regime. In future work, we may study the phase space of our cavity with three scatterers to get a better idea of the origin of these localized modes; however, these highly absorbing modes are not those which can lead to RSMs with tuning or optimization, as they are farthest from the real axis in the overdamped regime.

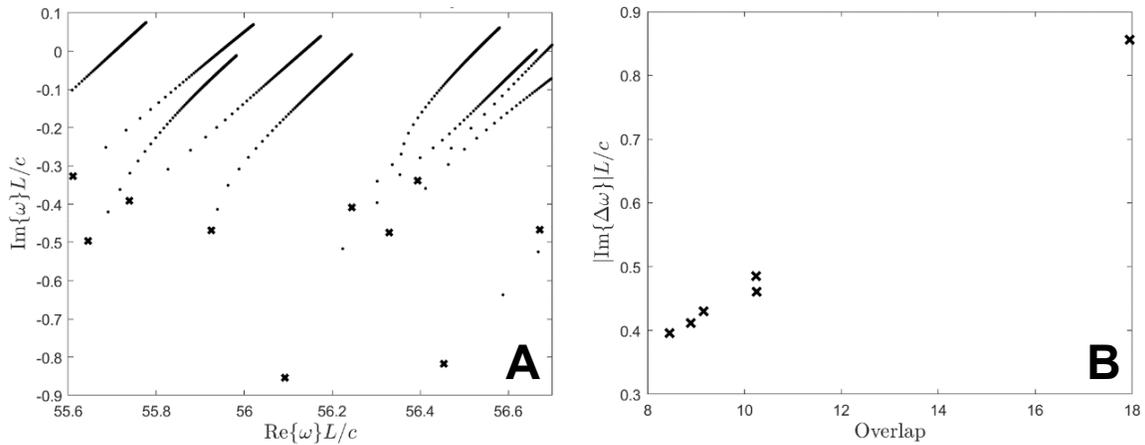

**Fig. S9.** (**A**) Tracked motion of the zeros of the scattering matrix in the complex plane as the wall conductivity is increased. The markers correspond to the lowest value of the conductivity. (**B**) Scatter plot showing the relation between the field overlap $\mathcal{O}$ with the cavity walls and the total vertical displacement of the zeros in the complex plane.

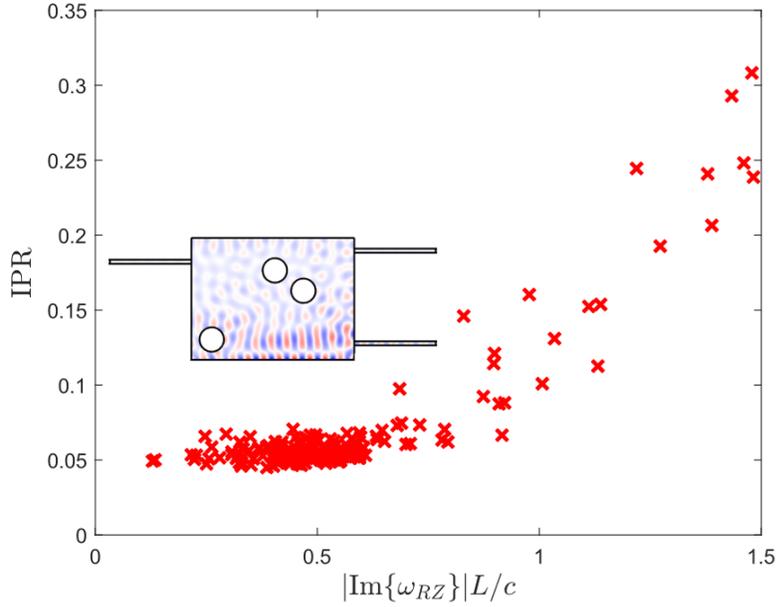

**Fig. S10.** Scatter plot of the IPR vs. the magnitude of the imaginary part of zeros of the scattering matrix. Larger IPR corresponds to more spatially localized modes and correlates with the magnitude of the imaginary part of the frequency. Inset: An example showing the localized nature of $\psi$ for a zero with $\text{IPR} = 0.25$ (the real part of $\psi$ is shown).

*Eigenvalue Interactions.* – The observation of differential absorption provides a partial explanation for the contrast in vertical distances covered by the different R-zeros, but it does not take into account any interaction between the zeros. The complex frequencies of the zeros are eigenvalues of a non-Hermitian operator and in general should repel each other in the complex plane. For systems which have scattering loss only (no absorption), it is known that, as outcoupling increases, these interactions get stronger and lead to a broadening of the distribution of resonances (the zeros are just complex conjugates in this case and their distribution also broadens). In general, some resonances actually move upwards as the scattering loss increases; this phenomenon is known in the literature as resonance trapping (*11, 12*). To the best of our knowledge, all research on this phenomenon was in relation to varying the openness of the system, and no systematic study exists on the presence or absence of resonance trapping when increasing the absorption in the cavity/system as an alternative loss channel. The zero motion shown in Fig. S9A does show hints of anti-crossings and repulsion of zeros, but no dramatic sign of some zeros moving upwards in analogy to resonance trapping models. While we do expect that the increase of absorption enhances eigenvalue repulsion and plays a role in the broadening of the distribution in both directions, further work is needed to confirm and quantify these effects. One difference with respect to existing work on resonance trapping that may play a role here is that absorption loss is uniform on the walls of our cavity, whereas scattering loss is typically localized and easier to avoid.

***RSM Observation with Single Continuously Tunable Parameter in Overdamped Regime***

In this section, we provide details of how we tuned our simulated cavity to the RSM shown in the inset of Fig. 1D in the main text. R-zeros are complex eigenvalues which are guaranteed by the analyticity of the wave equation to evolve continuously with a continuous external perturbation. Hence, once an R-zero with small imaginary part is found, continuously varying one suitable tuning parameter should be sufficient to move the R-zero arbitrarily close to the real axis, even in the overdamped regime. In our experiments, each programmable meta-atom is only 1-bit programmable, such that we cannot continuously tune one of our system's programmable parameters to observe how an R-zero moves arbitrarily close to the real-frequency axis and becomes an RSM. In contrast, in our simulations, this is feasible.

The continuously tunable parameter is the position of the lead on the left side of the cavity. We choose $\sigma = 60\,\epsilon_0 \omega_c$ for the conductivity of the cavity walls such that the average absorption is 97.6 %, where $\omega_c$ is the central angular frequency of the spectrum shown in the inset of Fig. 1D. We obtain this average absorption value by evaluating the quantity $\langle 1 - |S_{11}|^2 - |S_{21}|^2 - |S_{31}|^2 \rangle$, where the average is taken across the same frequency spectrum. We are hence in the strongly overdamped regime.

The curves in the inset of Fig. 1D look qualitatively similar to our experimental results (e.g., Fig. 1E) and the RSM is also similarly close to the real frequency axis. These observations give us confidence that our simulations faithfully represent the essential physics of our experimental strongly overdamped scattering system.

We note that the continuous tuning of a single parameter discussed in this section allows us to observe an RSM, albeit at an uncontrolled real frequency. A constraint on the real frequency at which the RSM occurs would yield the sort of optimization problem tackled in our experiments (see discussion of Fig. 1E and Fig. 1F).

*RSM Functionalization in Unitary Scattering System*

In the main text, we demonstrated multiple in-situ reprogrammable reflectionless signal routers that functionalized RSMs in a massively parametrized scattering system *with strong absorption*. In this section, we seek to demonstrate that the functionalization of RSMs is also possible in *unitary* scattering systems. While creating such functionalized RSMs is intuitively plausible in the lossless case, it has not to our knowledge been shown elsewhere. Ultimately, for applications of functionalized RSMs, the goal is to create such RSMs for demultiplexing and other routing functions in low-loss systems. Specifically, we consider the lossless case of our simulation setup (zero wall absorption), and we optimize the location of the three scatterers inside the cavity in order to implement a reflectionless wavelength demultiplexer (without constraints on the two demultiplexed frequencies).

The PML method for calculating R-zeros can be used as a tool to search for devices with different functionalities, similarly to what we do in our experiments. We demonstrate this in the following by looking for a structure that functions as a demultiplexer. An RSM is a necessary but not a sufficient condition for a demultiplexer; hence, in our algorithm, we start with an eigenvalue solver that finds the R-zeros in a rectangular region in the complex plane. R-zeros with small imaginary part correspond to dips in the reflection spectrum, so we next perform scattering calculations at the real part of these R-zeros to determine how much of the flux goes into each port. In the optimization, our objective is to maximize simultaneously the flux at one frequency into one of the two possible output ports, and the flux at another frequency into the other possible output port. Unlike our experiments, we chose not to impose stringent constraints on the two frequencies that are to be demultiplexed. Combining these two objectives, we aim to maximize the following function:

$$\mathcal{F} = \max_{\omega_{\text{RZ}}}\bigl(|S_{21}(\text{Re}\{\omega_{\text{RZ}}\})|^2)\bigr) \max_{\omega_{\text{RZ}}}\bigl(|S_{31}(\text{Re}\{\omega_{\text{RZ}}\})|^2)\bigr),$$

where the maximum is taken over the R-zeros that we obtain from the PML method for a given configuration of the scatterers. $|S_{21}|^2$ and $|S_{31}|^2$ are the transmissions into ports 2 and 3, respectively. Because it is very unlikely that the same $\omega_{\text{RZ}}$ yields the largest transmission into both ports simultaneously, we do not have to include an explicit constraint that the $\omega_{\text{RZ}}$ in the first and second term of the definition of $\mathcal{F}$ should be different.

We use a simple genetic algorithm for our optimization, with our search space corresponding to the location of the three scatterers. To simplify the problem, and to avoid that our algorithm would consider scenarios in which different scatterers overlap, we divide the space inside the cavity into $4 \times 6$ grid points. This discretization turns our continuous search space into a discrete one with $\binom{24}{3} = 2024$ distinct configurations. Despite using a small population size ($N_\text{pop} = 10$) and a relatively small number of generations ($N_\text{gen} = 40$), the genetic algorithm converges to a relatively good reflectionless demultiplexer. Figure S11 shows for the optimized cavity configuration both the reflection at the input port, as well as the two transmissions into the two output ports. The desired transmissions reach 98 % and 99 %.

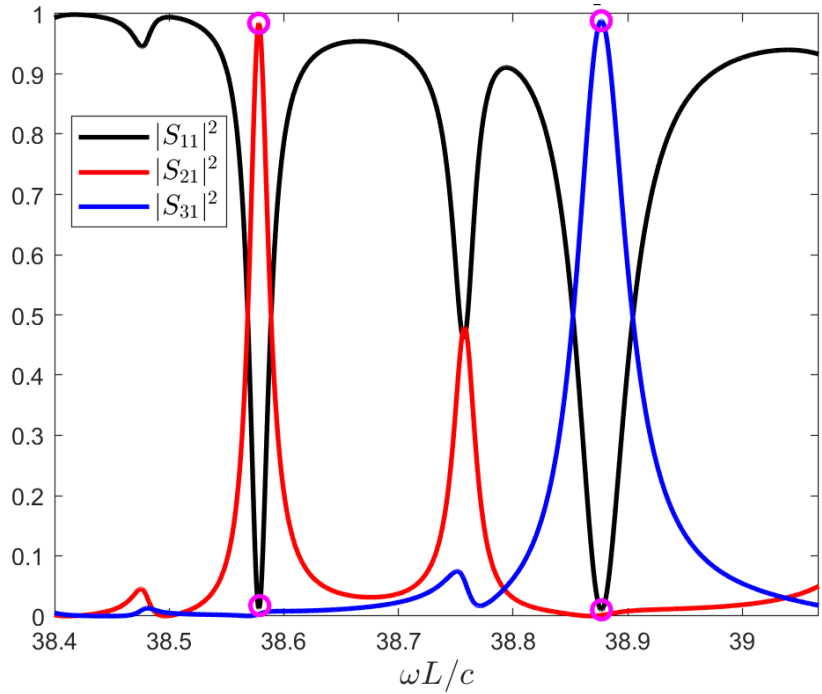

**Fig. S11.** Demonstration of a reflectionless demultiplexer in a unitary scattering system (no absorption). The optimized positions of the three scatterers are $(-0.2375, 0.2625)\,L$, $(0.0792, -0.0875)\,L$ and $(-0.2375, -0.0875)\,L$. The two demultiplexed angular frequencies are $38.5781\,c/L$ and $38.8769\,c/L$ (marked with circles). The transmission into port 2 at the first frequency is 98.42 %; the transmission into port 3 at the second frequency is 98.83 %. The small unwanted scattering is mainly reflection.

# Supplementary References